\documentclass[12pt]{article}
\usepackage{amsthm,amsmath,amssymb}
\usepackage{mathrsfs,psfrag,graphicx}

\newcommand{\QQ}{\tilde{\sol}_{\freq}}

\newcommand{\be}{b}
\newcommand{\dt}{\mathrm{d}_t}
\newcommand{\lite}{\kappa}
\newcommand{\Espace}[1][1]{\mathscr{H}_{#1,r}}

\newcommand{\ap}{a_{\parallel}}
\newcommand{\ab}{a_{\bot}}

\newcommand{\Anrm}[1]{|#1|_{\infty}}

\newcommand{\Veff}{V_{\mathrm{eff}}}
\newcommand{\iu}{{\rm i}}

\newcommand{\sol}{\eta}
\newcommand{\solw}{\sol_\freq}
\newcommand{\solp}{\sol_\Par}

\newcommand{\RIII}[1]{R_{\sol}^{(3)}(#1)}
\newcommand{\NII}[1]{N_{\sol}(#1)}

\newcommand{\dotp}[2]{\langle #1,#2\rangle}

\newcommand{\freq}{\mu}

\newcommand{\mom}{p}

\newcommand{\set}[1]{\mathrm{#1}}

\newcommand{\Hess}{\mathop{\mathrm{Hess}}}

\newcommand{\lexp}[1]{\mathrm{e}^{#1}}
\newcommand{\RE}{\mathop{\set{Re}}}
\newcommand{\IM}{\mathop{\set{Im}}}

\newcommand{\diag}{\mathop{\mathrm{diag}}}
\newcommand{\sign}{\mathop{\mathrm{sign}}}

\newcommand{\Null}[1]{\set{N}( #1 )}

\newcommand{\adj}{*}
\newcommand{\adjoint}[1]{#1^{\adj}}

\newcommand{\Eq}{Eq.\/~}

\newcommand{\Eref}[1]{\Eq\eqref{#1}}

\newcommand{\ie}{{\it i.e.\/}, }

\newcommand{\eg}{{\it e.g.\/}, }
\newcommand{\diff}{\mathop{\mathrm{\mathstrut{d}}}\!}

\newcommand{\Hone}{\set{H}_1}

\newcommand{\Sob}[1]{\set{H}_{#1}}
\newcommand{\C}[1]{\set{C}^{#1}}
\newcommand{\Ltwo}{\set{L}^2}
\newcommand{\Lp}[1]{\set{L}^{#1}}
\newcommand{\sind}[1]{{\text{{\tiny $#1$}}}}
\newcommand{\eps}{\epsilon_{\sind{V}}}
\newcommand{\epsE}{\epsilon_{\sind{0}}}
\newcommand{\epsH}{\epsilon_{\sind{h}}}
\newcommand{\epsT}{\epsilon}

\newtheorem{proposition}{Proposition}[section]

\newtheorem*{lemma*}{Lemma}
\newtheorem{remark}{Remark}[section]

\newtheorem{theorem}{Theorem}[section]
\newtheorem{lemma}[proposition]{Lemma}
\newtheorem{corollary}[proposition]{Corollary}

\newcommand{\Pn}{\mathcal{P}}

\newcommand{\En}{\mathcal{E}}
\newcommand{\Nn}{\mathcal{N}}
\newcommand{\Hn}{\mathcal{H}}

\newcommand{\HV}{\Hn_V}
\newcommand{\HnV}{\Hn_{V=0}}
\newcommand{\Ew}{\En_{\freq}}

\newcommand{\DE}{\Ew'}
\newcommand{\Mf}{\set{M}_{\mathrm{s}}}
\newcommand{\TM}{\set{T}_{\sol}\Mf}
\newcommand{\Span}{\mathop{\mathrm{span}}}
\newcommand{\Oh}{\mathcal{O}}
\newcommand{\nrmHo}[1]{\|#1\|_{\Hone}}
\newcommand{\Enrm}[1]{\|#1\|_{\Espace}}

\newcommand{\Laplace}{\Delta}

\newcommand{\zvec}{z}
\newcommand{\za}{\zvec_{\mathrm{t}}}
\newcommand{\zg}{\zvec_{\mathrm{g}}}
\newcommand{\zt}{\zvec_{\mathrm{b}}}
\newcommand{\zn}{\zvec_{\mathrm{s}}}

\newcommand{\tr}{\za}
\newcommand{\gu}{\zg}
\newcommand{\ga}{\zt}
\newcommand{\sa}{\zn}

\newcommand{\ptr}{\alpha^{\mathrm{tr}}}
\newcommand{\pbo}{\alpha^{\mathrm{b}}}

\newcommand{\nrmFree}[1]{\|#1\|}
\newcommand{\nrm}[1]{\|#1\|_{\Ltwo}}
\newcommand{\Bnrm}[1]{\big\|#1\big\|_{\Ltwo}}

\newcommand{\TB}{\mathcal{S}_{a \mom \gamma}}
\newcommand{\TBn}{\mathcal{S}_{a_0 \mom_0 \gamma_0}}
\newcommand{\TBf}{\mathcal{S}_{a(t) \mom(t) \gamma(t)}}

\newcommand{\VR}{\mathcal{R}_{V}}

\newcommand{\Rot}{\mathcal{T}^{\textrm{g}}_{\gamma}}
\newcommand{\Tr}{\mathcal{T}_a^{\textrm{tr}}}
\newcommand{\Bo}{\mathcal{T}_{\mom}^{\textrm{boost}}}

\newcommand{\Ph}{\mathcal{T}_{\mom}^{\textrm{b}}}

\newcommand{\Par}{\sigma}
\newcommand{\ParZ}{\tilde{\Par}_0}
\newcommand{\aZ}{\tilde{a}_0}
\newcommand{\mZ}{\tilde{\mom}_0}
\newcommand{\gZ}{\tilde{\gamma}_0}
\newcommand{\fZ}{\tilde{\freq}_0}
\newcommand{\wZ}{w_0}
\newcommand{\solZ}{\sol_{\ParZ}}
\newcommand{\mapPar}{\varsigma}
\newcommand{\spar}{\alpha}
\newcommand{\pars}{\underline{\spar}}
\newcommand{\gen}{\mathcal{K}}
\newcommand{\sgen}{\underline{\gen}}
\newcommand{\LL}{\mathcal{L}_\sol}

\newcommand{\lag}{\spar\cdot\gen}
\newcommand{\sag}{\pars\cdot\sgen}

\newcommand{\kx}{\langle x\rangle}
\newcommand{\kex}{\langle \eps x \rangle}
\newcommand{\kax}[1]{\langle #1 \rangle}

\makeatletter
\def\fixNumberingInArticle{
\@addtoreset{figure}{section}
\@addtoreset{equation}{section}
\renewcommand{\thefigure}{\thesection.\arabic{figure}}  
\renewcommand{\theequation}{\thesection.\arabic{equation}}  
}
\makeatother

\title{Long Time Motion of NLS Solitary
  Waves in a Confining Potential}

\author{B. L. G. Jonsson$^{1,3,}$\footnote{Supported by the Swiss National Foundation under NF-Project 20-105493 and by `The Year of PDE' at The Fields Institute.}, J. Fr\"ohlich$^1$, S. Gustafson$^{2,}$\footnote{Supported by NSERC 22R80976.},\\ I. M. Sigal$^{4,5,}$\footnote{Supported by NSF under Grant DMS-0400526.}\\$^1${\small Institute f\"{u}r Theoretische Physik, ETH H\"{o}nggerberg, Z\"{u}rich, Switzerland.} \\
$^2${\small Department of Mathematics, University of British Columbia, 
Vancouver, Canada.} \\
$^3${\small The Alfv\'enlaboratory, Royal Institute of 
Technology, Stockholm, Sweden.} \\
$^4${\small Department of Mathematics, University of Notre Dame, 
Notre Dame, IN, USA.} \\
$^5${\small Department of Mathematics, University of Toronto, Toronto, Canada.
}}

\begin{document}
\fixNumberingInArticle
\maketitle

\begin{abstract}
  We study the motion of solitary-wave solutions of a family of
  focusing generalized nonlinear Schr\"odinger equations with a
  confining, slowly varying external potential, $V(x)$.
  
  A Lyapunov-Schmidt decomposition of the solution combined with
  energy estimates allows us to control the motion of the solitary wave over a
  long, but finite, time interval.
 
  We show that the center of mass of the solitary wave
  follows a trajectory close to that of a Newtonian
  point particle in the external potential $V(x)$ over a long 
  time interval.
\end{abstract}

\section{Introduction}
\label{sec:intro}

We consider a family of generalized nonlinear Schr\"odinger and 
Hartree equations with a focusing nonlinearity. These equations have
solitary wave solutions, and, in this paper, we study the effective
dynamics of such solitary waves. The equations have the form:
\begin{equation}\label{eq:NLS}
 \iu \partial_t \psi(x,t) = -\Laplace \psi(x,t) + V(x)\psi(x,t) - f(\psi)(x,t),
\end{equation}
where $t\in \mathbb{R}$ is time, $x\in\mathbb{R}^d$ denotes a point in
physical space, $\psi:\mathbb{R}^d\times\mathbb{R}\mapsto \mathbb{C}$
is a (one-particle) wave function, $V$ is the external potential,
which is a real-valued, confining, and slowly varying function on
$\mathbb{R}^d$, and $f(\psi)$ describes a nonlinear self-interaction
with the properties that $f(\psi)$ is ``differentiable'' in $\psi$,
$f(0)=0$, and $f(\bar{\psi})=\overline{f(\psi)}$.  Precise assumptions
on $V$ and $f$ are formulated in Section~\ref{sec:ass}.

The family of nonlinearities of interest to us 
includes local nonlinearities, such as
\begin{equation}
f(\psi)=\lambda|\psi|^{s}\psi, \ 0<s<\frac{4}{d},\ \lambda>0,
\end{equation}
and Hartree nonlinearities
\begin{equation}
f(\psi)=\lambda(\Phi*|\psi|^{2})\psi, \ \lambda>0,
\end{equation}
where the (two-body) potential $\Phi$ is real-valued, of positive
type, continuous, spherically symmetric, and tends to 0 as
$|x|\rightarrow \infty$. Here $\Phi*g:=\int \Phi(x-y)g(y)\diff^d y$ denotes
convolution.  Such equations are encountered in the theory of Bose
gases (BEC), in nonlinear optics, in the theory of water waves and in
other areas of physics.

It is well known that \Eref{eq:NLS} has solitary wave solutions when
$V\equiv 0$. 
Let $\solw\in\Ltwo$ be a
spherically symmetric, positive solution of the nonlinear
eigenvalue problem
\begin{equation}\label{eq:1sol}
-\Laplace \sol + \freq\sol-f(\sol)=0.
\end{equation}
The function $\solw$ is called a ``solitary wave profile''.
Among the solitary wave solutions of \eqref{eq:NLS} are Galilei transformations of $\solw$, 
\begin{equation}\label{eq:solW}
\psi_{\text{sol}}:=\mathcal{S}_{a(t)p(t)\gamma(t)}\sol_{\freq(t)},
\end{equation}
where $\TB$ is defined by
\begin{equation}\label{eq:TB0}
(\TB\psi)(x):=\lexp{\iu\mom\cdot(x-a)+\iu \gamma}\psi(x-a). 
\end{equation}
Let $\Par:=\{a,\mom,\gamma,\freq\}$, where $\freq$ is as in
\Eref{eq:1sol}. For $\psi_{\text{sol}}$ to be a solution to
\eqref{eq:NLS} with $V\equiv 0$ the modulation parameters, $\Par$,
must satisfy the equations of motion
\begin{equation}\label{eq:mod}
a(t)=2p t+a,\ \mom(t)=\mom,\ \gamma(t)=\freq t +\mom^2 t+\gamma, \
\freq(t)=\freq
\end{equation}
with $\gamma\in \mathbb{S}^1$, $a,\mom\in\mathbb{R}^d$, $\freq\in
\mathbb{R}_+$. In other words, 
$\Par$ satisfy \eqref{eq:mod}, then
\begin{equation}
\psi_{\text{sol}}(x,t)=(\mathcal{S}_{a(t)p(t)\gamma(t)}\sol_{\freq(t)})(x)
\end{equation}
solves \Eref{eq:NLS} with $V\equiv 0$.  Thus \eqref{eq:solW}, with
$a(t),\mom(t),\gamma(t),\freq(t)$ as above, describes a
$2d+2$-dimensional family solutions of \Eref{eq:NLS} with $V\equiv 0$.
Let the {\bf soliton manifold}, $\Mf$ be defined by
\begin{equation}
   \Mf := \{\TB\sol_{\freq}
                : \{a,\mom , \gamma,\freq \} \in 
   \mathbb{R}^d\times 
   \mathbb{R}^d\times \mathbb{S}^1 \times I \} \; ,
\end{equation}
where $I$ is a bounded interval in $\mathbb{R}_+$.

Solutions to \eqref{eq:1sol} behave roughly like 
$\lexp{-\sqrt{\freq}|x|}$, as $|x|\rightarrow
\infty$. So $\sqrt{\freq}$ is a reciprocal length
scale that indicates the ``size'' of the solitary wave.

\begin{figure}[htbp]
\psfrag{M}{$\Mf$} 
\psfrag{P}{$\psi(\cdot,t)$}
\psfrag{s}{$\sol_{\Par(t)}$} 
   \centering
   \centerline{\includegraphics{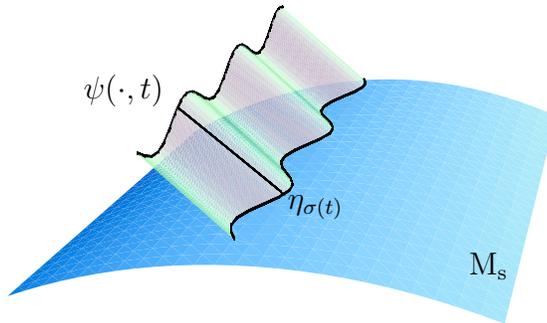}}
   \parbox{\linewidth}{
   \caption{The trajectory $\psi(\cdot,t)$ over the soliton Manifold $\Mf$.}
\label{fig:a}}
\end{figure}

We consider the Cauchy problem for \Eref{eq:NLS}, with initial
condition $\psi_0$ in a weighted Sobolev space. For Hartree
nonlinearities, global wellposedness is known \cite{Enno}. For local
nonlinearities, the situation is more delicate; see
Condition~\ref{con:GWP} and Remark~\ref{rem:GWP} in
Section~\ref{sec:ass}.  Let the initial condition $\psi_0$ be
``close'' to $\Mf$. Then, we will show, the corresponding solution 
$\psi$ will remain ``close'' to $\Mf$, over a long time interval.  
A certain ``symplectically orthogonal'' projection of $\psi$ onto $\Mf$ is then
well defined and traces out a unique curve on $\Mf$.  We denote this
curve by $\sol_{\Par(t)}$, see Figure~\ref{fig:a}.

An essential part of this paper is to
determine the leading order behavior of 
$\Par(t) = \{a(t),p(t),\gamma(t),\mu(t)\}$ and to estimate error
terms. To this end, let $W$ be a smooth, positive,
polynomially bounded function, and define
\begin{equation}\label{eq:18}
V(x)=W(\eps x)
\end{equation}
where $\eps$ is a small parameter. Furthermore, let $\psi_0$ be an
initial condition ``$\epsE$--close'' to $\sol_{\Par_0}\in\Mf$, for some
$\Par_0$. Roughly speaking, this initial condition has 
length scale $1/\sqrt{\freq}_0$.  We
will consider external potentials, $V$, as in \eqref{eq:18}, for a scaling
parameter $\eps$ satisfying 
\begin{equation}
\eps \ll \sqrt{\freq_0},
\end{equation}
\ie we assume that the external potential varies very little over the
length scale of $\psi_0$. For simplicity, we choose $\freq=1$ and $\eps\ll 1$,
at the price of re-scaling the nonlinearity.

We decompose the solution $\psi$ of
\eqref{eq:NLS} into a part which is a solitary wave and a small part,
a ``perturbation'', $w$.  That is, we write $\psi$ as
\begin{equation}
\psi = \TB(\solw+w).
\end{equation}
This does not define a unique decomposition, unless $2d+2$ additional
conditions are imposed. These conditions say that the perturbation $w$
is `symplectically orthogonal' to the soliton manifold $\Mf$. 

The main idea used to control the perturbation $w$ is to derive
differential equations in time for the modulation parameters, $\Par$,
which depend on the external potential. These equations appear
naturally when one projects solutions of \eqref{eq:NLS} onto the
soliton manifold. To control the motions of $\Par$ and $w$, we make
use of conserved quantities: the energy
\begin{equation}
\HV(\psi):=\frac{1}{2}\int (|\nabla \psi|^2 + V|\psi|^2) 
\diff^d x - F(\psi),
\end{equation}
where $F'(\psi)=f(\psi)$ (this is a variational derivative), 
the mass (or charge)
\begin{equation}
\Nn(\psi):=\frac{1}{2}\int |\psi|^2\diff^d x,
\end{equation}
and the ``almost conserved'' momentum
\begin{equation}
\Pn(\psi):=\frac{1}{4}\int (\bar{\psi} \nabla \psi 
- \psi\nabla \bar{\psi})\diff^d x.
\end{equation}
To achieve control over the perturbation $w$, we introduce a `Lyapunov
functional' 
\begin{equation}
\Lambda(\psi,t):=K_{\Par}(\psi)-K_{\Par}(\TB\solw),
\end{equation}
where
$\Par=\Par(t) = \{ a(t), p(t), \gamma(t), \mu(t) \}$, and where
\begin{equation}
\begin{split}
K_{\Par}(\psi) &:= \HV(\psi)+(\mom^2+\freq)\Nn(\psi)-2\mom\cdot \Pn(\psi) \\
&-\frac{1}{2}\int\big( V(a)+\nabla V(a)\cdot (x-a)\big)
|\psi|^2\diff^d x,
\end{split}
\end{equation}
\ie $K_{\Par}$ is essentially a linear combination of the conserved
and almost conserved quantities.  Using the linear transformation
$u:=\TB^{-1}\psi$, we change questions about the size of fluctuations
around $\TB\solw$ to ones about the size of fluctuations around the
solitary wave profile $\sol_{\freq(t)}$.
In this ``moving frame'', the $K_{\Par}(\psi)$ terms in the Lyapunov functional
introduced above take the form
\begin{equation}
K_{\Par}(\TB u)=\Ew(u)+\frac{1}{2}\int \VR|u|^2\diff^d x,
\end{equation}
where 
\begin{equation}
\VR(x):= V(x+a)-V(a)-\nabla V(a)\cdot (x-a)
\end{equation}
and
\begin{equation}\label{eq:Ew}
\Ew(u):=\HnV(u)+\freq \Nn(u).
\end{equation}
In the moving frame the Lyapunov functional depends on the parameters
$\freq$ and $a$, but not on $\mom$ and $\gamma$.  Furthermore, $\solw$
is a critical point of $\Ew(\solw)$, \ie $\Ew'(\solw)=0$. The change
of frame discussed above simplifies the analysis leading to our main
result. 

Simply stated, our main theorem shows that, for initial
conditions $\psi_0$ $\epsE$-close to $\Mf$, the perturbation $w$ is
of order $\epsT:= \eps+\epsE$, for all times smaller than $C\epsT^{-1}$.
Furthermore, the center of mass of the solitary wave, $a$, and the
center of mass momentum $\mom$ satisfy the following equations
\begin{eqnarray}
\dot a = 2p + \Oh(\epsT^2), && \dot p = - \nabla V(a)+\Oh(\epsT^2).
\end{eqnarray}
The remaining modulation parameters $\freq$ and 
$\gamma$ satisfy
\begin{eqnarray}
\dot \freq =\Oh(\epsT^2), && \dot\gamma = \freq-V(a)+p^2+\Oh(\epsT^2).
\end{eqnarray}
A precise statement is found in the next section. 

This is the first result of its type covering
{\it confining} external potentials.
Indeed, we can exploit the confining nature of the potential
to obtain a {\it stronger} result than that of \cite{FGJS-I}
(and that stated above) for a certain class of initial conditions
which we now describe.  Consider the classical
Hamiltonian function: 
\begin{equation}
  h(a,p):=\big(p^2+V(a)\big)/2.
\end{equation}  
Given an initial
condition $\psi_0$ $\epsE$--close to $\sol_{\Par_0}\in \Mf$, where
$\Par_0=\{a_0,p_0,\gamma_0,\freq_0\}$, we require the initial
position $a_0$ and momentum $p_0$ to satisfy
\begin{equation}
h(a_0,p_0) - \min_{a} h(a,0) \leq  \epsH,
\end{equation} 
with $\eps\leq C\epsH\leq 1$, for some constant $C$. For this class of
initial conditions, our main result shows that the perturbation $w$
remains $\Oh(\epsT)$ for longer times:
\begin{equation}
\label{eq:longer}
t<\frac{C}{\eps\sqrt{\epsH}+\epsT^2}.
\end{equation}
This improvement is non-trivial. For example, it means that 
we can control the perturbation of a solitary wave which undergoes
many oscillations near the bottom of a potential well. 

\noindent{\bf Remark:}
  We can also extend our analysis to a class of slowly time-dependent
  external potentials without much additional work.  
  We introduce a scale parameter,
  $\tau$, in time: $V(x,t):=W(\eps x,\tau t)$. To 
  determine the size of $\tau$ heuristically we consider
\begin{equation}
\frac{d}{dt}h(a,\mom,t) = \mom\big(\dot \mom +\nabla V(a,t)\big)
+ \frac{1}{2}(\dot a-2p)\cdot \nabla V(a,t) + \partial_t V(a,t).
\end{equation}
We want the last two terms to have the same size.  The second
but last term is of size $\epsT^2\eps$, since $\dot a$ satisfy the
classical equations of motion to order $\epsT^2$. The last term is of
size $\tau$. Thus if $\tau$ is chosen to be $\tau=\Oh(\eps^3)$ 
all our estimates will survive.

The following example suggests that accelerating solitary wave solutions of 
\Eref{eq:NLS} in a confining external potential can, in fact, survive for  
arbitrarily long times. Choose
$V(x):= x \cdot A x + d \cdot x + c \geq 0$ and $A>0$ (positive matrix).  
Then \eqref{eq:NLS} has the following solution:
\begin{equation}\label{eq:124}
\psi(x,t)=\lexp{\iu p(t)\cdot(x-a(t))+\iu \gamma(t))} 
\tilde{\eta}_\freq(x-a(t))
\end{equation}
with 
\begin{equation}\label{eq:125}
\dot p = -\nabla V(a), \ \dot a=2p,\ \dot \gamma=p^2+\freq-V(a),
\end{equation}
where $\tilde{\eta}_\freq$ solves the equation
\begin{equation}
-\Laplace \eta + \freq \eta - f(\eta) + (x \cdot A x) \eta = 0.
\end{equation}
Thus, given a solution of the equations of motion \eqref{eq:125}, a
family of solitary wave solutions is given by \eqref{eq:124}, for arbitrary
times $t$.  For details see Appendix \ref{sec:fam}.

The first results of the above type, for bounded, time-independent
potentials were proved in \cite{Frohlich+Tsai+Yau2000,
Frohlich+Tsai+Yau2002} for the Hartree equation under a spectral
assumption. This result was later extended to a general class of
nonlinearities in \cite{FGJS-I}.  
Neither of these works deals with a confining external potential.
In particular, their results do not extend to the longer
time interval~(\ref{eq:longer}) described above.

For local pure-power nonlinearities
and a small parameter $\eps$, it has been shown in 
\cite{Bronski+Jerrard2000} that if an initial condition is
of the form $\TBn\sol_{\freq_0}$, then the solution $\psi(x,t)$ of
\Eq\eqref{eq:NLS} satisfies
\begin{equation}
\eps^{-d}|\psi(\frac{x}{\eps},\frac{t}{\eps})|^2
\rightarrow \nrm{\solw}^2 \delta_{a(t)}
\end{equation}
in the $\C{1*}$ topology (dual to $\C{1}$), provided $a(t)$ satisfies
the equation $\frac{1}{2}\ddot{a}=\nabla W(a)$, where $V(x)=W(\eps
x)$. This result was strengthened in \cite{Keraani2002} for a bounded
external potential and in \cite{Carles2003} for a potential given by a
quadratic polynomial in $x$. 

There have been many recent works on asymptotic properties for generalized
nonlinear Schr\"odinger equations. Asymptotic stability, scattering
and asymptotic completeness of solitary waves for bounded external
potential tending to 0 at $\infty$ has been shown under various
assumptions. See for example, \cite{Soffer+Weinstein1988,
  Soffer+Weinstein1990,Soffer+Weinstein1992, BP92,
  Buslaev+Perelman1995,Cuccagna2001,Cuccagna2002,
  Buslaev+Sulem2002,
  Tsai+Yau2002,Tsai+Yau2002b,Tsai+Yau2002c,RSS,Soffer+Weinstein2004,
  Gustafson+Nakanishi+Tsai2004,Gang+Sigal2004,Perelman2004}.

Though these are all-time results, where ours is long (but finite)-time,
our approach has some advantages: we can handle confining potentials
(for which the above-described results are meaningless); we require a much
less stringent (and verifiable) spectral condition; we track the 
finite-dimensional soliton dynamics (Newton equations); and our
methods are comparatively elementary.

Our paper is organized as follows. In Section~\ref{sec:ass}, we state
our hypotheses and the main result. In Section~\ref{sec:2}, we recall
the Hamiltonian nature of \Eref{eq:NLS} and describe symmetries of
\eqref{eq:NLS} for $V\equiv 0$. We give a precise definition of the
soliton manifold $\Mf$ and its tangent space.  In Section~\ref{sec:3},
we introduce a convenient parametrization of functions in a small
neighborhood of $\Mf$ in phase space, and we derive equations for the
modulation parameters $\Par=\{a,\mom,\gamma,\freq\}$ and the
perturbation $w$ around a solitary wave $\sol_{\Par}=\TB\solw$.  In
this parametrization, the perturbation $w$ is symplectically
orthogonal to the tangent space $\set{T}_{\solp}\Mf$ to $\Mf$ at
$\solp$. In Section \ref{sec:rel}, we similarly decompose the initial
condition $\psi_0$ deriving in this way the initial conditions,
$\Par_0$ and $w_0$, for $\Par$ and $w$, and estimating $w_0$.  In
Section \ref{sec:5}, we derive bounds on the solitary wave position,
$a$, and the momentum, $\mom$, by using the fact that the Hamiltonian,
$h(a,p)$ is almost conserved in time. In Section \ref{sec:4}, we
construct the Lyapunov functional, $\Lambda(\psi,t)$, and compute its time
derivative.  This computation is used in Section~\ref{sec:6} in order
to obtain an upper bound on $\Lambda(\psi,t)$. This bound, together with the
more difficult lower bound derived in Section~\ref{sec:7}, is used in
Section~\ref{sec:end} in order to estimate the perturbation $w$ and
complete the proof of our main result, Theorem~\ref{thm:main}.  Some
basic inequalities are collected in
Appendices~\ref{app:RVbd}--\ref{sec:ene}.  In Appendix~\ref{sec:fam},
we construct a family of time-dependent solutions with parameters
exactly satisfying the classical equations of motion.

\section{Notation, assumptions and main result}
\label{sec:ass}

Let $\Lp{s}$ denote the usual Lebesgue space of functions,
$\C{s}$ the space of functions with $s$ continuous
derivatives, and $\Sob{s}$ the Sobolev space of order $s$.  
Abbreviate $\kx^2:=1+|x|^2$.

\paragraph{Assumptions on the external potential.} 
Let $W(x)$ be a $\C{3}$ function, and let $\min_x W(x)=0$. 
Let $\beta\in\mathbb{Z}^d$ with $\beta_j\geq 0$
$\forall j=1,\ldots,d$ be a multi-index. Given a number $r\geq 1$
let $W$ be such that
\begin{eqnarray}\label{eq:Wup}
&|\partial_x^\beta W(x)|\leq C_{\max{V}}
\kx^{r-|\beta|} \ \text{for} \ |\beta|\leq 3,
&  \\
\label{eq:Wlow} 
&\Hess W(x) \geq \rho_1 \kx^{r-2},&
\end{eqnarray}
and 
\begin{equation}\label{eq:Wfar}
W(x)\geq c_V|x|^r,\ \text{for}\ |x|\geq c_L
\end{equation}
for some positive constants $C_{\max{V}}$, $\rho_1$, $c_V$, $c_L$. 

The number $r$ is called the growth rate of the external potential.
Here $\Hess W$ is the Hessian of $W$ with respect to spatial
variables.  Define $V(x):=W(\eps x)$. Then, for $r\geq 1$,
\begin{eqnarray}\label{eq:Vup}
& |\partial_x^\beta V(x)|\leq C_V \eps^{|\beta|} \kex^{r-|\beta|},\
\text{for} \ |\beta|\leq 3, &
\\ \label{eq:Vlow}
& \Hess V(x) \geq \rho_1 \eps^2 \kex^{r-2}, &
\end{eqnarray}
 and
\begin{equation}\label{eq:Vfar}
V(x)\geq c_V(\eps|x|)^r,\ \text{for}\ \eps|x|\geq c_L.
\end{equation}

\paragraph{Assumptions on the initial condition $\psi_0$.}  
The energy space, $\Espace$, for a given growth rate $r$ of the
external potential, is defined as
\begin{equation}\label{eq:Espace}
\Espace:=\{\psi\in \Hone:\langle x\rangle^{r/2}\psi \in \Lp{2}\}.
\end{equation}
Let $\Espace'$ denote the dual space of $\Espace$.
The energy norm is defined as
\begin{equation}
\Enrm{\psi}^2:=\nrmHo{\psi}^2+\nrm{\kex^{r/2}\psi}^2
\end{equation}
We require $\psi_0\in \Espace$.

\medskip

In what follows, we identify complex functions 
with real two-component functions
via 
\[
  \mathbb{C} \ni \psi(x) = \psi_1(x) + \iu \psi_2(x) \;
  \longleftrightarrow \; \vec{\psi}(x) = (\psi_1(x), \psi_2(x)) 
  \in \mathbb{R}^2.
\]
Consider a real function $F(\vec{\psi})$ on a space
of real two-component functions, and let
$F'(\vec{\psi})$ denote its $L^2$-gradient.
We identify this gradient with a complex function
denoted by $F'(\psi)$.  Then
\[
  F'(\bar{\psi}) = \overline{F'(\psi)} \;
  \longleftrightarrow \;
  F(\sigma \vec{\psi}) = F(\vec{\psi}), 
\]
where $\sigma :=\diag(1,-1)$, since the latter property 
is equivalent to 
$F'(\vec{\psi}) = \sigma F'(\sigma \vec{\psi})$.

\paragraph{Assumptions on the nonlinearity $f$.}
\begin{enumerate}
\item\label{con:GWP} (GWP \cite{Cazenave1996,Yajima+Zhang2001,Yajima+Zhang2004,Enno}) Equation~\eqref{eq:NLS} is globally
  well-posed in the space $\set{C}(\mathbb{R},\Espace) \cap
  \C{1}(\mathbb{R},\Espace')$.  See Remark~\ref{rem:GWP} below.

\item\label{con}
The nonlinearity $f$ maps from $\Hone$ to $\Sob{-1}$, with $f(0)=0$.
$f(\psi)=F'(\psi)$ is the $\Ltwo$-gradient  
of a $C^3$ functional $F : H_1 \to \mathbb{R}$ 
defined on the space of real-valued,
two-component functions, satisfying
the following conditions:
\begin{enumerate}
\item (Bounds)
\label{con:A}
\begin{equation}\label{eq:Taylor}
\sup_{\nrmFree{u}_{\Hone}\leq M}
\nrmFree{F''(u)}_{\set{B(\Hone,\Sob{-1})}}<\infty, \
\sup_{\nrmFree{u}_{\Hone}\leq M}\ \nrmFree{F'''(u)}_{\Hone \mapsto
\set{B}(\Hone,\Sob{-1})}<\infty,
\end{equation}
where $\set{B}(X,Y)$ denotes the space of bounded linear operators from $X$ to $Y$.
\item(Symmetries \cite{FGJS-I})
\label{con:sym}
  $F(\mathcal{T}\psi)=F(\psi)$ where $\mathcal{T}$ is either translation
  $\psi(x)\mapsto \psi(x+a)$ $\forall a\in\mathbb{R}^d$, or spatial
  rotation $\psi(x)\mapsto \psi(R^{-1}x)$, $\forall R\in \set{SO}(d)$, 
  or boosts $\Ph: u(x)\mapsto \lexp{\iu \mom\cdot x}u(x)$, $\forall
  \mom\in\mathbb{R}^d$, 
  or gauge transformations 
  $\psi\mapsto \lexp{\iu \gamma}\psi$, $\forall \gamma\in
  \mathbb{S}^1$, or complex conjugation $\psi \mapsto \bar{\psi}$.
\end{enumerate}
\item\label{con:F} (Solitary waves) \label{con:Sol} There exists a
  bounded open interval $\tilde{I}$ on the positive real axis such
  that for all $\freq\in \tilde{I}$:
\begin{enumerate} 
\item (Ground state 
\cite{Berestycki+Lions+Peletier1981,Berestycki+LionsI1983,Berestycki+LionsII1983,McLeod1993})
The equation
\begin{equation}
\label{eq:gs}
-\Laplace \psi + \freq \psi - f(\psi)=0.
\end{equation}
has a spherically symmetric, positive $\Ltwo\cap \C{2}$ solution,
$\sol = \sol_\mu$.
\item\label{con:stab} (Stability: see \eg
  \cite{Grillakis+Shatah+Strauss1990}) This solution, $\sol$, satisfies
\begin{equation}
\partial_\freq \int \sol^2_\mu \diff^d x>0.
\end{equation}
\item\label{con:Null} (Null space condition: see \eg \cite{FGJS-I})
Let $\LL$ be the linear operator
\begin{equation}
        \LL:=\begin{pmatrix} L_1 & 0 \\ 0 & L_2 \end{pmatrix}
\end{equation}
where $L_1:=-\Laplace+\freq -f^{(1)}(\sol)$, and $L_2:=-\Laplace +
\freq - f^{(2)}(\eta)$, with
$f^{(1)}:=\Big(\partial_{\RE{\psi}}\big(\RE(f)\big)\Big)(\sol)$, 
and $f^{(2)}:=\Big(\partial_{\IM{\psi}}\big(\IM(f)\big)\Big)(\sol)$.
We require that
\begin{equation}
\Null{\LL}=\Span\{\begin{pmatrix}0 \\ \sol \end{pmatrix}, 
\begin{pmatrix} \partial_{x_j}\sol\\ 0\end{pmatrix},\ j=1,\ldots, d\}.
\end{equation}
\end{enumerate}
\end{enumerate}
Conditions~\ref{con}--\ref{con:F} on the nonlinearity are discussed
in \cite{FGJS-I}, where further references can be found. Examples of
nonlinearities that satisfy the above requirements are local
nonlinearities
\begin{equation}\label{eq:n1}
f(\psi)=\beta |\psi|^{s_1}\psi + \lambda |\psi|^{s_2}\psi, \ 0<s_1<s_2<\frac{4}{d},\ \beta\in \mathbb{R},\ \lambda>0,
\end{equation}
and Hartree nonlinearities
\begin{equation}\label{eq:n2}
f(\psi)=\lambda(\Phi*|\psi|^2)\psi, \ \lambda>0,
\end{equation}
where $\Phi$ is of positive type, continuous and spherically symmetric and
tends to 0, as $|x|\rightarrow \infty$.
Of course, $\lambda$ can be scaled out by rescaling
$\psi$. For precise conditions on $\Phi$ we refer to
\cite{Cazenave1996,Enno}.  
\begin{remark}\label{rem:GWP}
  For Hartree nonlinearities global well-posedness is known for
  potentials $0\leq V\in \Lp{1}_{loc}$ \cite{Enno}. For local
  nonlinearities, the situation is more delicate. Global
  well-posedness and energy conservation is known for potentials with
  growth-rate $r\leq 2$~\cite{Cazenave1996}. For $r>2$ and local
  nonlinearities, local well-posedness has been shown in
  the energy space~\cite{Yajima+Zhang2001,Yajima+Zhang2004}. 
  For local nonlinearities, a proof of the energy conservation
  needed for global well-posedness, and the
  application of this theory to our results, is missing.
\end{remark}
For $V\equiv 0$, \Eref{eq:NLS} is the usual generalized nonlinear
Schr\"odinger (or Hartree) equation. For self-focusing nonlinearities
as in examples \eqref{eq:n1} and \eqref{eq:n2}, it has stable solitary
wave solutions of the form
\begin{equation}\label{eq:solp}
\sol_{\Par(t)}(x):=\lexp{\iu \mom(t)\cdot (x-a(t))+\iu \gamma(t)}\sol_{\freq(t)}(x-a(t)),
\end{equation}
where $\sigma(t):=\{a(t),\mom(t),\gamma(t),\freq(t)\}$, and
\begin{equation}
a(t)=2pt+a,\ \gamma(t)=\freq t + p^2 t+ \gamma,\
p(t)=p,\ \freq(t)=\freq,
\end{equation} 
with $\gamma\in\mathbb{S}^1$, $a,p\in \mathbb{R}^d$ and $\freq\in
\mathbb{R}^+$, and where $\solw$ is the spherically symmetric,
positive solution of the nonlinear eigenvalue problem
\begin{equation}\label{eq:sol}
-\Laplace \sol + \freq \sol - f(\sol)=0.
\end{equation}
Recall from \eqref{eq:TB0} that the linear map $\TB$ is defined as 
\begin{equation}\label{eq:TB}
(\TB g)(x):= \lexp{\iu \mom\cdot (x-a)+\iu \gamma}g(x-a).
\end{equation}

In analyzing solitary wave solutions to \eqref{eq:NLS} we encounter
two length scales: the size $\propto \freq^{-1/2}$ of the support of
the function $\solw$, which is determined by our choice of initial
condition $\psi_0$, and a length scale determined by the potential, $V$,
measured by the small parameter $\eps$. We consider the regime,
\begin{equation}
\frac{\eps}{\sqrt{\freq}}\ll 1.
\end{equation}

We claim in the introduction that if $\psi_0$ is close to $\solp$, for
some $\Par$ then we retain control for times $\propto\epsT^{-1}$.
Restricting the initial condition to a smaller class of $\solp$, with
small initial energy, we retain control for longer times.  In our
main theorem, which proves this claim, we wish to treat both
cases uniformly. To this end, let $\epsH$ and $K$ be positive numbers
such that $\epsH\in K[\eps,\min_{\freq\in I}\sqrt{\freq}]$ 
and assume
\begin{equation}
h(a_0,\mom_0):=\frac{1}{2}\big(\mom_0^2+V(a)\big)\leq \epsH
\end{equation}
(recall $\min_a V(a)=0$).  The lower bound for $\epsH$
corresponds to our restricted class of initial data, 
the upper bound to the larger class of data. 
In particular, $\epsH\geq K \eps$.

We are now ready to state our main result.
Fix an open proper sub-interval $I \subset \tilde{I}$.
\begin{theorem}\label{thm:main}
Let $f$ and $V$ satisfy the conditions listed above. 
There exists $T>0$ such that for 
$\epsilon:=\eps+\epsE$ sufficiently small, 
and $\epsH\geq K\eps$,
if the initial condition $\psi_0$ satisfies
\begin{equation}\label{eq:ebnd}
\nrmHo{\psi_0-\TBn\sol_{\freq_0}} +
\nrm{\kex^{r/2}(\psi_0-\TBn\sol_{\freq_0})}\leq \epsE
\end{equation}
for some $\sigma_0:=\{a_0,\mom_0,\gamma_0,\freq_0\}\in
\mathbb{R}^d\times \mathbb{R}^d\times\mathbb{S}^1\times I$ such that
\begin{equation}\label{eq:hbound}
h(a_0,p_0)\leq \epsH,
\end{equation}
then for times $0\leq t\leq T(\eps\sqrt{\epsH}+\epsT^2)^{-1}$, 
the solution to \Eref{eq:NLS}
with this initial condition is of the form
\begin{equation}
\psi(x,t)=\TBf\big(\sol_{\freq(t)}(x)+w(x,t)\big),
\end{equation}
where
$\nrmHo{w}+\Bnrm{\kex^{r/2}w}\leq C\epsT$. The modulation
parameters $a,\mom,\gamma$ and $\freq$ satisfy the differential
equations
\begin{align}\label{eq:thp}
\dot p& = -( \nabla V)(a) + \Oh(\epsT^2), \\
\dot a& = 2p + \Oh(\epsT^2), \\
\dot \gamma& = \freq -V(a)+\mom^2+\Oh(\epsT^2), \\
\dot \freq &= \Oh(\epsT^2).\label{eq:thf} 
\end{align}
\end{theorem}

\begin{remark}[Remark about notation]
  Fr\'echet derivatives are always understood to be defined on real
  spaces. They are denoted by primes. $C$ and $c$ denote various
  constants that often change between consecutive lines and
  which do not depend on $\eps$, $\epsE$ or $\epsT$.
\end{remark}

\section{Soliton manifold}
\label{sec:2}
In this section we recall the Hamiltonian nature of \Eref{eq:NLS} and
some of its symmetries.  We also define the soliton manifold and its
tangent space.

An important part in our approach is played by the variational
character of \eqref{eq:NLS}. More precisely, the nonlinear
Schr\"odinger equation \eqref{eq:NLS} is a Hamiltonian
system with Hamiltonian
\begin{equation}\label{eq:HV}
\HV(\psi) := \frac{1}{2}\int (|\nabla \psi|^2 + V|\psi|^2)\diff^d x - F(\psi).
\end{equation}
The Hamiltonian $\HV$ is conserved  \ie
\begin{equation}
\HV(\psi)=\HV(\psi_0).
\end{equation}
A proof of this can be found, for local nonlinearities and $r\leq 2$,
in \eg Cazenave~\cite{Cazenave1996}, and for Hartree nonlinearities in
\cite{Enno}.  An important role is played by the mass
\begin{equation}\label{eq:N}
\Nn(\psi):= \int |\psi|^2 \diff^d x,
\end{equation}
which also is conserved,
\begin{equation}
\Nn(\psi(t))=\Nn(\psi_0).
\end{equation}

We often identify complex spaces, such as the Sobolev space
$\Hone(\mathbb{R}^d,\mathbb{C})$, with real spaces; \eg
$\Hone(\mathbb{R}^d,\mathbb{R}^2)$, 
using the identification
$\psi=\psi_1+\iu\psi_2 \leftrightarrow (\psi_1,\psi_2)=:\vec{\psi}$.
With this identification, the complex structure $\iu^{-1}$ corresponds to
the operator
\begin{equation}
J:=\begin{pmatrix}0 & 1 \\ -1 & 0\end{pmatrix}.
\end{equation}
The real $\Ltwo$-inner product in the real notation is
\begin{equation}
\dotp{\vec u}{\vec w} := \int (u_1w_1 + u_2w_2) \diff^d x,
\end{equation}
where $\vec{u}:=(u_1,u_2)$.
In the complex notation it becomes
\begin{equation}
\dotp{u}{w} := \RE \int u\bar{w} \diff^d x.
\end{equation}
We henceforth abuse notation and drop the arrows.  The symplectic form is
\begin{equation}
\omega(u,w):=\IM\int u\bar w \diff^d x.
\end{equation}
We note that $\omega(u,w)=\dotp{u}{J^{-1}v}$ in the real notation.

Equation~\eqref{eq:NLS} with $V\equiv 0$ is invariant
under spatial translations, $\Tr$, gauge transformations, $\Rot$, and
boost transformations, $\Bo$, where
\begin{equation}
    \Tr :\psi(x,t) \mapsto \psi(x-a,t) \; , \ 
    \Rot : \psi(x,t)\mapsto \lexp{\iu\gamma}\psi(x,t) 
        \label{eq:T1} \; , 
\end{equation}
\begin{equation}
    \Bo: \psi(x,t)\mapsto 
    \lexp{\iu(\mom\cdot x - \mom^2 t) }
    \psi(x-2 \mom t,t)\; .
        \label{eq:T2}
\end{equation}
The transformations~\eqref{eq:T1}--\eqref{eq:T2} 
map solutions of eq.~\eqref{eq:NLS} with $V\equiv 0$ 
into solutions of \eqref{eq:NLS} with $V\equiv 0$.

Let $\Ph:\psi(x)\mapsto \lexp{\iu\mom\cdot x}\psi(x)$ be the $t=0$
slice of the boost transform.  The combined symmetry transformations
$\TB$ introduced in \eqref{eq:TB} can be expressed as
\begin{align}\label{eq:Sym}
\TB\sol = \Tr\Ph\Rot \solw(x)
=\lexp{\iu (\mom \cdot (x-a)+\gamma)}\solw(x-a).
\end{align}

We define the soliton manifold as
\begin{equation}
   \Mf := \{\TB\sol_{\freq}
                : \{a,\mom , \gamma,\freq \} \in 
   \mathbb{R}^d\times 
   \mathbb{R}^d\times \mathbb{S}^1 \times I \} \; .
\end{equation}
The tangent space to this manifold at the solitary wave 
profile $\solw\in \Mf$ is given by
\begin{equation}
   \set{T}_{\solw}\Mf = \Span(\za,\zg,\zt,\zn) \; ,
\end{equation}
where
\begin{align}
   \tr :=&\left.\nabla_a  \Tr \solw \right|_{a=0} = \begin{pmatrix} -\nabla \solw\\ 0 \end{pmatrix}\; ,
   \label{eq:t1} && 
   \gu := \left. \frac{\partial}{\partial \gamma} 
        \Rot \solw\right|_{\gamma=0} = \begin{pmatrix} 0 \\ \solw\end{pmatrix} \; ,\\
    \ga := &\left. \nabla_\mom \Bo \solw \right|_{\mom=0,t=0} = 
        \begin{pmatrix} 0 \\ x\solw \end{pmatrix}\;,  &&  \label{eq:t4}
   \sa :=  \begin{pmatrix} \partial_{\freq} \solw \\ 0 \end{pmatrix}\; .
\end{align}
Above, we have explicitly written the basis of tangent vectors in the real
space. 

Recall that the equation~\eqref{eq:gs} can be written as
$\Ew'(\sol_\mu) = 0$ where
\[
  \Ew(\psi) = \mathcal{H}_{V \equiv 0}(\psi) + \frac{\mu}{2} \Nn(\psi).
\]
Then the tangent vectors listed above are generalized zero modes of
the operator $\mathcal{L}_\mu := \Ew''(\sol_\mu)$.  That is,
$(J\mathcal{L}_\mu)^2 z = 0$ 
for each tangent vector $z$ above.  
To see this fact for $\gu$, for example, 
recall that $\DE(\psi)$ is gauge-invariant.
Hence $\DE(\Rot \solw)=0$. Taking the derivative with respect to the parameter
$\gamma$ at $\gamma=0$ gives $\LL \gu=0$. 
The other relations are derived analogously (see \cite{Weinstein1985}).

\section{Symplectically orthogonal decomposition}
\label{sec:3}

In this section we make a change of coordinates for the Hamiltonian
system $\psi\mapsto (\Par,w)$, where $\Par:=(a,\mom,\gamma,\freq)$. 
We also give the equations in this new set of coordinates. 

Let
\begin{equation}\label{eq:m}
m(\freq):=\frac{1}{2}\int \sol_\freq^2(x)\diff^d x.
\end{equation}
Let
\begin{equation}\label{eq:CI}
C_I:=\max_{\substack{z\in \{x\solw,\solw,\nabla\solw,\partial_\freq \solw\}\\ \freq\in \tilde{I}}}
(\nrmHo{z},\nrm{\kex^{r/2}z},\nrm{\gen z}).
\end{equation}
When it will not cause confusion, 
for $\Par = \{ a, p, \gamma, \mu \}$ we will abbreviate
\[
  \sol_{\Par} := \TB \eta_\mu.
\]

Now define the neighborhood of $\Mf$:
\begin{equation}\label{eq:Udelta}
  U_\delta := \{\psi\in L^2 :\inf_{\Par\in \Sigma}
  \nrm{\psi-\sol_{\Par}}\leq \delta\},
\end{equation}
where $\Sigma := \{a,\mom,\gamma,\freq: a\in \mathbb{R}^d,
\mom\in \mathbb{R}^d,\gamma\in \mathbb{S}^1, \freq \in I\}$. 
Our goal is to decompose a given function $\psi\in U_\delta$
into a solitary wave and a perturbation:
\begin{equation}\label{eq:splitt}
\psi = \TB(\solw + w).
\end{equation}
We do this according to the following theorem.
Let  $\tilde{\Sigma} := \{a,\mom,\gamma,\freq: a\in \mathbb{R}^d,
\mom\in \mathbb{R}^d,\gamma\in \mathbb{S}^1, \freq \in \tilde{I}\}$.
\begin{theorem}\label{thm:splitt}
  There exists $\delta > 0$ and a unique 
  map $\mapPar\in \C{1}(U_\delta,\tilde{\Sigma})$ such that 
  (i)
\begin{equation}
\dotp{\psi-\sol_{\mapPar(\psi)}}{J^{-1} z}=0, \;\; 
\forall z\in \set{T}_{\sol_{\mapPar(\psi)}}\Mf, \;\;
\forall \psi \in U_\delta
\end{equation}
and (ii) if, in addition, 
$\delta \ll (2C_I)^{-1}\min(m(\freq),m'(\freq))$ 
then there exists a constant
$c_I$ independent of $\delta$ such that
\begin{equation}\label{eq:Omega}
\sup_{\psi\in U_\delta} \nrm{\varsigma'(\psi)}\leq c_I.
\end{equation}
\end{theorem}
\begin{proof}
  Part (i): Let the map $G:L^2 \times \tilde{\Sigma} \mapsto \mathbb{R}^{2d+2}$
  be defined by
\begin{equation}
G_j(\psi,\varsigma):=\dotp{\psi-\sol_{\varsigma}}{J^{-1}z_{\varsigma,j}},
\ \forall j=1,\ldots 2d+2.
\end{equation}
Part (i) is proved by applying the implicit function theorem 
to the equation $G(\psi,\varsigma)=0$, around a
point $(\sol_{\Par},\Par)$. For details we refer to Proposition~5.1 in
\cite{FGJS-I}.

Part (ii): Abbreviate:
\begin{equation}
\Omega_{jk}:=\dotp{\partial_{\varsigma_j}\sol_{\varsigma}}
{J^{-1}z_{\varsigma,k}},
\end{equation}
where $z_{\varsigma,k}$ is the $k$:th element of
$\TB\{\tr,\gu,\ga,\sa\}$.  By explicitly inserting the tangent
vectors, we find that $\nrm{\Omega} \geq \inf_{\freq\in
  I}(m(\freq),m'(\freq))$.  Thus, $\Omega$ is invertible by Condition
\ref{con:stab} in Section~\ref{sec:ass}.

From a variation of $\psi$ in $G(\psi,\varsigma(\psi))=0$ we find
\begin{equation}
\varsigma'_k(\psi)=\sum_{j=1}^{2d+2} 
(J^{-1}z_{\varsigma})_j(\tilde{\Omega}^{-1})_{jk}.
\end{equation}
where 
\begin{equation}
\tilde{\Omega}_{jk}:=\Omega_{jk} + 
\dotp{\psi-\sol_{\varsigma(\psi)}}{J^{-1}\partial_{\varsigma_j}z_{\varsigma,k}}
\end{equation}
Using the upper bound of $\delta$, and the definition of $C_I$ above, we find
\begin{equation}
\sup_{\psi\in U_\delta} \nrm{\varsigma'(\psi)}\leq 
\frac{2C_I}{\inf_{\freq\in I}(m(\freq),m'(\freq))}=:c_{I}. 
\end{equation}
\end{proof}

We now assume $\psi(t) \in U_\delta\cap\Espace$, 
and set $\Par(t):=\varsigma(\psi(t))$ as defined by
Theorem~\ref{thm:splitt}.
Write
\begin{equation}\label{eq:udef}
  u:=\TB^{-1}\psi = \sol_\mu + w
\end{equation}
so that $w$ satisfies
\begin{equation}
\dotp{w}{J^{-1} z}=0, \;\; 
\forall z\in \set{T}_{\sol_{\mu}}\Mf.
\end{equation}
Here $u$ is the solution in a moving frame.

Denote the anti-self-adjoint infinitesimal generators of symmetries as
\begin{equation}
   \gen_{j} = \partial_{x_j}, \ \  
   \gen_{d+j} = \iu x_j, \ \   \gen_{2d+1}=\iu ,\ \
   \gen_{2d+2}=\partial_\freq,\ \ j=1,...,d 
   \label{eq:gen}
\end{equation}
and define corresponding coefficients 
\begin{equation}
   \alpha_{j} = \dot{a}_j - 2\mom_j, \ \ 
   \alpha_{d+j} = -\dot{\mom}_j - \partial_{x_j} V(a), \ \  
   j=1,...,d,     \label{eq:mu} 
\end{equation}
\begin{equation}
   \alpha_{2d+1} =
\freq-\mom^2+\dot{a}\cdot \mom
        -V(a)-\dot{\gamma},
\ \ \alpha_{2d+2}  = -\dot{\freq}. \label{eq:nu}
\end{equation}
Denote
\begin{equation}
\sag := \sum_{j=1}^{2d+1} \alpha_j \gen_j, \ \ \text{and}\ \ 
\lag := \sag + \alpha_{2d+2}\partial_\freq.
\end{equation}

Substituting $\psi=\TB u$ into \eqref{eq:NLS} we obtain 
\begin{equation}\label{eq:dut}
\iu \dot u = \DE(u) + \VR u +\iu \sag u,
\end{equation}
where 
\begin{equation}\label{eq:VR}
\VR(x) = V(x+a) - V(a) - \nabla V(a) \cdot x.
\end{equation}
To obtain the equations for $(\Par,w)$ we project Eqn.~\eqref{eq:dut}
onto $\TM$ and $(J\TM)^{\bot}$ and use \eqref{eq:udef}. We illustrate this
method of deriving the equations for $\Par$, for the
projection of \eqref{eq:dut} along $\iu \sol$:
\begin{equation}\label{eq:proj1}
\dotp{\sol}{\dot\freq\partial_\freq\sol+\dot w}=\dotp{\iu \sol}{\LL w+
\NII{w}+\VR (\sol+w)+\iu \sag (\sol+w)}.
\end{equation}
where we have used $u=\sol+w$ and $\DE(u)=\LL w+\NII{w}$
where $\LL := \Ew''(\sol)$ is given explicitly as
\begin{equation}\label{eq:LL}
\LL w = -\Laplace w +\freq w - f'(\sol)w.
\end{equation}
In particular, for local nonlinearities of the form $g(|\psi|^2)\psi$, we 
have in the complex notation, since $\sol(x)\in \mathbb{R}$,
\begin{equation}
\LL w := -\Laplace w +\freq w - g(\sol^2)w - 2\sol g'(\sol^2)\RE w.
\end{equation}
Here
\begin{equation}\label{eq:NII}
\NII{w} := - f(\sol + w ) + f(\sol) + f'(\sol)w.
\end{equation}
We find the equation for $\dot\freq$ once we note that
$\partial_t\dotp{\sol}{w}=0$, $\LL\iu\sol=0$, $\dotp{\iu
  \sol}{\VR\sol}=0$, $\dotp{\sol}{\sgen\sol}=0$ and
$\adjoint{\sgen}=-\sgen$. Inserting this into \eqref{eq:proj1} gives
\begin{equation}
\dot\freq m'(\freq) = 
\dotp{\iu \sol}{\NII{w}+\VR w}-\alpha\cdot\dotp{\gen\sol}{w}.
\end{equation}
The projection along the other directions works the same way:
we use the fact that these directions are the generalized
zero modes of $\LL$, and furthermore that they are orthogonal to $Jw$. The
calculations are worked out in detail in \cite{FGJS-I} (See
Eqns.~(6.20)--(6.22) in \cite{FGJS-I}.) We give the result:
\begin{align}
\dot{\gamma}
& = \freq-\mom^2+\dot{a}\cdot \mom- V(a)- 
(m'(\freq))^{-1}\left(\dotp{\partial_\freq \sol}{\NII{w}+ \VR w}  
\right. \label{eq:gam} \\ & \left. \quad 
- \alpha\cdot\dotp{\mathcal{K} \partial_\freq \sol}{\iu w}  
 + \dotp{\partial_\freq \sol}{\VR \sol}
\right),\nonumber
\\ \nonumber \\
\dot{\freq}&=\big(m'(\freq)\big)^{-1} \left(
\dotp{\iu\sol}{\NII{w}+\VR w} - \alpha\cdot\dotp{\mathcal{K}
\sol}{w}\right),  \label{eq:dotmu}
\end{align}
\begin{align}
\dot{a}_k&=2\mom_k+ \big(m(\freq)^{-1}\big)\left(\dotp{\iu x_k \sol}{\NII{w}+ \VR
w}-\alpha\cdot\dotp{\mathcal{K} x_k \sol}{w}
\right) , \label{eq:tr}
\\ \nonumber \\
\dot{\mom}_k & = -\partial_{a_k} V(a) + (m(\freq))^{-1}\big(-\frac{1}{2}\dotp{(\partial_{x_k}\VR)\sol}{\sol}+
\dotp{\partial_k\sol}{\NII{w}+\VR w}\nonumber  \\ & \quad  - 
\alpha\cdot\dotp{\mathcal{K} \partial_k \sol}{\iu w} \big), \label{eq:bo}
\end{align} 
and
\begin{equation}\label{eq:dw}
\iu \dot{w} = \LL w + N(w) + \VR(\sol+w) + \iu \sag(\sol+w) - \iu
\dot{\freq} \partial_\freq \sol.
\end{equation}

Note that the first two terms on the right-hand side of 
Eqn.~\eqref{eq:bo} can be written as $-\partial_{a_k}\Veff(a,\freq)$, where
\begin{equation}\label{eq:Veff}
\Veff(a,\freq):= \nrm{\solw}^{-2}
\int V(a+x)|\solw(x)|^2 \diff^d x.
\end{equation}
Hence,
\begin{equation}
\dot \mom_k = -\nabla_a \Veff(a,\freq) + (m(\freq)^{-1}
\dotp{\partial_{x_k}\solw}{\NII{w}} + \Oh(\nrm{w}(\eps^2+|\alpha|)),
\end{equation}
where $|\alpha|^2=\sum |\alpha_j|^2$.

Thus we have obtained the dynamical equations for 
$(\Par,w)$. 
\begin{remark}
The transformation 
\begin{equation}
\Par:=(a,p,\gamma,\freq)\mapsto\hat{\Par}:= (a,P,\gamma,m)
\end{equation} with $P:=\frac{1}{2}p\nrm{\solw}^2$ and
$m:=\frac{1}{2}\nrm{\solw}^2$ gives a canonical symplectic structure
and Darboux coordinates on $\Mf$, \ie for $w=0$
\begin{align}
\dot P &= -\partial_a \HV(\TB\solw), &&
\dot a = \partial_P \HV(\TB\solw), \\
\dot m & = \partial_\gamma \HV(\TB\solw), &&
\dot \gamma  = -\partial_{m} \HV(\TB\solw).
\end{align}
Here $\nabla_{\hat{\Par}} \HV(\TB\solw) = (m\nabla_a
\Veff,2P/m,0,-P^2/m^2+V(a)-\freq)$. 
\end{remark}

\section{Initial conditions $\ParZ$, $\wZ$.}\label{sec:rel}

\begin{figure}[htbp]
\psfrag{a}{$\psi_0$} 
\psfrag{b}{$\sol_{\sigma_0}$}
\psfrag{c}{$\sol_{\varsigma(\psi_0)}=\solZ$} 
\psfrag{H}{$\Hone$}
\psfrag{M}{$\Mf$}
   \centering
   \centerline{\includegraphics{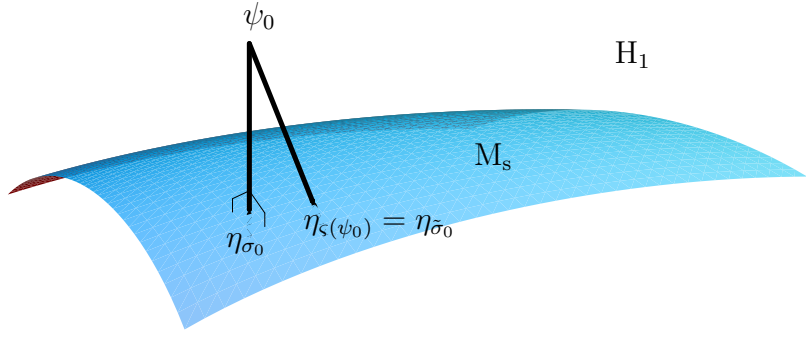}}
   \parbox{\linewidth}{
   \caption{Orthogonal decomposition versus skew-orthogonal decomposition.}\label{fig:1}}
\end{figure}

In this section we
use Theorem~\ref{thm:splitt} in order to decompose the initial condition $\psi_0$ as (see Figure~\ref{fig:1})
\begin{equation}
\psi_0 = \mathcal{S}_{\aZ,\mZ,\gZ}(\sol_{\fZ}+\wZ)
\end{equation}
so that $\wZ \bot J^{-1}\set{T}_{\sol_{\fZ}}\Mf$. This decomposition
provides the initial conditions $\ParZ$ and $\wZ$, for the parameters,
$\Par$, and fluctuation, $w$ (determined for later times by
Theorem~\ref{thm:splitt}). The main work here goes into estimating
$\wZ$.
Let $\varsigma:U_{\delta}\mapsto \tilde{\Sigma}$ 
be the map established in
Theorem~\ref{thm:splitt}. Then $\ParZ=\{\aZ,\mZ,\gZ,\fZ\}$ 
and $\wZ$ are given as
$\ParZ:=\varsigma(\psi_0)$ and 
\begin{equation}\label{eq:wZ}
\wZ:=\mathcal{S}_{\aZ\mZ\gZ}^{-1}
  (\psi_0-\solZ), \ \wZ \bot J\set{T}_{\sol_{\fZ}}\Mf.
\end{equation}
Recall the definitions of $\gen$ \eqref{eq:gen}, and $C_I$
\eqref{eq:CI}. Theorem~\ref{thm:splitt} states $\sup_{\psi\in
  U_\delta} \nrm{\varsigma'(\psi)} \leq c_{I}$.

Bounds for $\wZ$ and $\ParZ$ are stated in the following
proposition
\begin{proposition}\label{prop:wo}
  Let $\wZ$ be defined as above. Let
  $\Par_0:=\{a_0,\mom_0,\gamma_0,\freq_0\}$ and let $\psi_0$ satisfy
  $\|\psi_0-\sol_{\Par_0}\|_{L^2} \leq \delta$ (where $\delta$
  is from Theorem~\ref{thm:splitt}), and let
  $\psi_0\in\Espace$.  Then there exists positive constants $C_1$,
  $C_2$, such that
\begin{align}\label{eq:N2}
|\ParZ-\Par_0|&\leq c_{I}\nrm{\psi_0-\sol_{\Par_0}},\\
\nrmHo{\wZ}& \leq C_1(1+p_0^4+\nrm{\psi_0-\sol_{\Par_0}}^4)\nrmHo{\psi_0-\sol_{\Par_0}} \label{eq:stat1} 
\end{align}
and
\begin{multline}
\nrm{\kex^{r/2}\wZ}\leq 3^{r/2}\nrm{\kex^{r/2}(\psi_0-\sol_{\Par_0})}
\\+ C_2(1+|p_0|^2+\eps^r|a_0|^r + \nrm{\psi_0-\sol_{\Par_0}}^2+\eps^r\nrm{\psi_0-\sol_{\Par_0}}^r)\nrm{\psi_0-\sol_{\Par_0}}. \label{eq:stat2}
\end{multline}
where $C_1$ and $C_2$ depend only on $C_I$, $c_I$ and $r$, where $C_I$
is defined in \eqref{eq:CI} and $c_I$ in Theorem~\ref{thm:splitt}.
\end{proposition}
\begin{proof}
  First we consider inequality \eqref{eq:N2}. Abbreviate
  $\ParZ:=\varsigma(\psi_0)$ and analogously for the components
  $a,\mom,\gamma,\freq$ of $\varsigma$. Let
  $|\varsigma|^2:=\sum_{j=1}^{2d+2}|\varsigma_j|^2$. From
  Theorem~\ref{thm:splitt} we know that $\varsigma(\psi)$ is a
  $\C{1}$-map. Thus, for $j\in 1,...,2d+2$ and some $\theta_1\in[0,1]$
\begin{equation}
(\ParZ-\Par_0)_j=\dotp{\varsigma_j'(\theta_1 \psi_0+(1-\theta_1)\sol_{\Par_0})}{(\psi_0-\sol_{\Par_0})}.
\end{equation}
Since  $\sup_{\psi\in U_{\delta}}\nrmFree{\varsigma'(\psi)}\leq c_{I}$
the inequality \eqref{eq:N2} follows.

Consider inequality \eqref{eq:stat1} and rewrite $w(\cdot,0)=:\wZ$
from \eqref{eq:wZ} as
\begin{equation}\label{eq:woo}
\wZ=\mathcal{S}_{\aZ\mZ\gZ}^{-1}
(\psi_0-\sol_{\Par_0})
+\mathcal{S}_{\aZ\mZ\gZ}^{-1}(\sol_{\Par_0}-\solZ).
\end{equation}
To estimate this, we first estimate the linear operator $\TB^{-1}$:
\begin{equation}\label{eq:HoTB}
\nrmHo{\TB^{-1}\psi} \leq 2(1+|\mom|^2)^{1/2}\nrmHo{\psi}.
\end{equation}
The first term in \eqref{eq:woo} is in the appropriate form, for the second
term we recall that $\sol$ is a $\C{1}$-map. Thus for some
$\theta_2\in[0,1]$
\begin{equation}\label{eq:N1}
\solZ-\sol_{\Par_0}=\sum_{j=1}^{2d+2}\left.
(\ParZ-\Par_0)_j
\partial_{\Par_j}\sol_{\Par}\right|_{\Par=\theta_1\ParZ+(1-\theta_2)\Par_0}.
\end{equation}
To calculate the norm of this expression, note that
\begin{equation}\label{eq:N3}
\partial_\Par \solp = \TB z_{\freq,\mom}, \ \text{where}\ 
z_{\freq,\mom}:=\{\iu \mom\solw+\nabla \solw,\iu x\solw,\iu \solw,\partial_\freq \solw\}
\end{equation}
and $\nrmHo{z_{\freq,\mom}} \leq \sqrt{5}C_I(1+|p|^2)^{1/2}$.  Let
$n(\Par,\Par_0):=(\Par-\Par_0)\theta_2+\Par_0$, and define
$g^2:=1+|\mZ-p_0|^2+p_0^2$. The $\Hone$-norm of \eqref{eq:N1}, using
\eqref{eq:HoTB} and \eqref{eq:N3} is
\begin{equation}
\begin{split}
\label{eq:dHone}
\nrmHo{\solZ-\sol_{\Par_0}}
&\leq |\ParZ-\Par_0| \big.\nrmHo{\partial_{\Par}\sol_{\Par}}
\big|_{\Par=n(\ParZ,\Par_0)} \\
&\leq \left. 2\sqrt{5}C_I(1+|p|^2)\right|_{p=n(\mZ,\mom_0)}|\ParZ-\Par_0|
\leq 9C_Ig^2|\ParZ-\Par_0|.
\end{split}
\end{equation}
We now calculate the $\Hone$ norm of $\wZ$ (see \eqref{eq:woo}) 
using \eqref{eq:N2}, \eqref{eq:HoTB} with momentum 
$p=\mZ-p_0+p_0$ and \eqref{eq:dHone}. We find
\begin{equation}
\begin{split}
\nrmHo{\wZ} &\leq 2g(\nrmHo{\psi_0-\sol_{\Par_0}}+
\nrmHo{\solZ-\sol_{\Par_0}})\\ 
&\leq 2g\big(1+9C_Ic_{I}g^2\big)
\nrmHo{\psi_0-\sol_{\Par_0}}.
\end{split}
\end{equation}
The coefficient above is less then $cg^4 +C$, and 
$g^4\leq 3(1+c_{I}^4\nrm{\psi_0-\sol_{\Par_0}}^4+|p_0|^4)$. Inserting
and simplifying gives the inequality \eqref{eq:stat1}.

The quantity appearing in the third and last inequality 
\eqref{eq:stat2}, can be rewritten as
\begin{equation}\label{eq:N4}
\kex^{r/2}w_0 = \kex^{r/2}\mathcal{S}_{\aZ\mZ\gZ}^{-1}\big(
(\psi_0-\sol_{\Par_0})+(\sol_{\Par_0}-\solZ)\big).
\end{equation}
We begin our calculation of the norm of \eqref{eq:N4} by considering
the linear operator $\kex^{r/2}\TB$. We have 
\begin{equation}\label{eq:N5}
\kex^{r/2}\TB\psi  =\TB\kax{\eps(x-a)}^{r/2}\psi
\end{equation}
and $\nrm{\TB\psi}=\nrm{\psi}$. From Lemma~\ref{lem:maxmin} we obtain
\begin{equation}\label{eq:xS}
\begin{split}
\nrm{\kex^{r/2}\TB\psi} &\leq \nrm{\kax{\eps(x-(a-a_0)-a_0)}^{r/2}\psi}\\ 
&\leq 3^{\max(r/2,r-1)}\big(\nrm{\kex^{r/2}\psi}+g_2\nrm{\psi}\big),
\end{split}
\end{equation}
where $g_2:=(\eps|a-a_0|)^{r/2}+(\eps|a_0|)^{r/2})$. 
Using this we find the $\Ltwo$-norm of \eqref{eq:N4} to be 
\begin{multline}\label{eq:q}
\nrm{\kex^{r/2}\wZ}\leq 
C\big(\nrm{\kex^{r/2}(\psi_0-\sol_{\Par_0})} + g_2\nrm{\psi_0-\sol_{\Par_0}} 
\\ + \nrm{\kex^{r/2}(\solZ-\sol_{\Par_0})}
+g_2\nrm{\solZ-\sol_{\Par_0}}\big).
\end{multline}
The first and second term of the above expression is in an appropriate
form. We bound the third term by using \eqref{eq:N1}, \eqref{eq:N3} and \eqref{eq:N5} to get
\begin{equation}\label{eq:xsol}
\begin{split}
\nrm{\kex^{r/2}(\solZ-\sol_{\Par_0})}
&\leq |\ParZ-\Par_0|\left. \nrm{\kax{\eps(x-a)}^{r/2}z_{p,\freq}}\right|_{\Par=n(\ParZ,\Par_0)} \\
&\leq 3^{\max(r/2,r-1)}\sqrt{5}C_Ig(1+g_2)
|\ParZ-\Par_0|.
\end{split}
\end{equation}
The last term of \eqref{eq:q} is straight forward to bound:
\begin{equation}
\begin{split}
\nrm{\solZ-\sol_{\Par_0}} &\leq |\ParZ-\Par_0|\left.\nrm{\partial_{\Par}\sol_{\Par}}\right|_{\Par=n(\ParZ,\Par_0)} \\  
&\leq |\ParZ-\Par_0|\big.\nrm{z_{p,\freq}}\big|_{\substack{\mom=n(\mZ,\mom_0) 
\freq=n(\fZ,\freq_0)}}
\leq  \sqrt{5}C_Ig|\ParZ-\Par_0|.
\label{eq:solDiff}
\end{split}
\end{equation}
Inserting \eqref{eq:xsol} and \eqref{eq:solDiff} into \eqref{eq:q} gives
\begin{multline}
\nrm{\kex^{r/2}\wZ}\leq C\Big(\nrm{\kex^{r/2}(\psi_0-\sol_{\Par_0})} 
 \\ +\big(g_2+g(1+2g_2)\big)
\nrm{\psi_0-\sol_{\Par_0}}\Big),
\end{multline}
where $C$ depend only on $C_I$, $c_I$ and $r$.
We simplify this, by repeatedly using Cauchy's inequality and \eqref{eq:N2} on 
the expression
in front of the $\nrm{\psi_0-\sol_{\Par_0}}$-term, to obtain
\begin{multline}
\nrm{\kex^{r/2}\wZ}\leq C\Big(\nrm{\kex^{r/2}(\psi_0-\sol_{\Par_0})} 
+ \big(1+\nrm{\psi_0-\sol_{\Par_0}}^2\\ +\eps^r\nrm{\psi_0-\sol_{\Par_0}}^r + |p_2|^2+(\eps|a_0|)^r\big)
\nrm{\psi_0-\sol_{\Par_0}}\Big).
\end{multline}
This gives the third inequality of the proposition.
\end{proof}

Recall the initial energy bound \eqref{eq:ebnd}
\begin{equation}
\nrmHo{\psi_0-\sol_{\Par_0}}+\nrm{\kex^{r/2}(\psi_0-\sol_{\Par_0})}\leq \epsE,
\end{equation}
and the bound on the initial kinetic and potential energy for 
the solitary wave \eqref{eq:hbound}
\begin{equation}
\frac{1}{2}(p_0^2+V(a_0))\leq \epsH.
\end{equation}
We have the corollary
\begin{corollary}\label{cor:wo}
Let \eqref{eq:ebnd}, \eqref{eq:hbound} and 
\eqref{eq:Vup}--\eqref{eq:Vfar} hold
with $\epsE < \delta$. Then
\begin{equation}
|\ParZ-\Par_0|\leq c_{I}\epsE, \ 
\nrmHo{\wZ}\leq C_{1}\epsE,
\end{equation}
\begin{align}
\nrm{\kex^{r/2}\wZ}&\leq C_{2}\epsE
\end{align}
and
\begin{equation}\label{eq:hesta}
h(\aZ,\mZ)\leq C_3(\epsH+ \epsE^2+\eps\epsE),
\end{equation}
where $C_1$, $C_2$ and $C_3$ depend only on $c_L$, $c_V$
(\Eref{eq:Vfar}), $C_E := \max (\eps,\epsE,\epsH)$ and the constants
in Proposition~\ref{prop:wo}.
\end{corollary}
\begin{proof}
  Starting from Proposition~\ref{prop:wo} the first three inequalities
  follow directly through the energy bounds \eqref{eq:ebnd}, 
  \eqref{eq:hbound} together with the observation that either 
  $\eps|a_0|\leq c_L$ or $c_V(\eps |a_0|)^r\leq V(a_0)\leq 2\epsH$.
  We also use that $\epsH$, $\epsE$ and
  $\eps$ are all bounded by a constant $C_E$.

The last inequality follows from the fact that 
$h(a,p):=(p^2+V(a))/2$
is a $\C{1}$ function. For some $\theta\in[0,1]$
\begin{equation}
\begin{split}
h(a,p)-h(a_0,p_0) &= ((p-p_0)\theta+p_0)\cdot(p-p_0) \\ 
&+ \frac{1}{2} (a-a_0)\cdot \nabla V((a-a_0)\theta+a_0).
\end{split}
\end{equation}
Thus, using \eqref{eq:Vup}, and $\kax{x+y}^{r-1}\leq 
3^{\max(0,(r-3)/2)}\big(1+2^{(r-1)/2}(|x|^{r-1}+|y|^{r-1})\big)$ gives
\begin{multline}
|h(a,p)-h(a_0,p_0)|\leq C\Big(|p-p_0|^2+|p_0|^2 + 
\\ 
\eps^2|a-a_0|\big(1+|\eps(a-a_0)|^{r-1}+|\eps a_0|^{r-1}\big)\Big).
\end{multline}
With $p=\mZ$ and $a=\aZ$ above, and $|\ParZ-\Par_0|\leq c_{I}\epsE$,
$h(a_0,p_0)\leq \epsH$, \eqref{eq:hbound} and \eqref{eq:Vfar} we have 
have shown \eqref{eq:hesta}.
\end{proof}

\section{Bounds on soliton position and momentum}\label{sec:5}

In this section we use the bounded initial soliton energy,
Corollary~\ref{cor:wo}, to find upper bounds on 
position and momentum of the solitary wave. We express the norms first in
terms of $h(\aZ,\mZ)$ and the small parameters. In
Corollary~\ref{cor:apest} we state the final result, 
where the bounds are just
constants times the small parameters $\epsE$, $\epsH$ and $\eps$.

Recall (see \eqref{eq:Vup} and \eqref{eq:Vfar})  
that the potential $V$ is non-negative and satisfies 
the following upper and lower bounds:
\begin{equation}\label{eq:Vup_2}
   |\partial_x^\beta V|\leq C_V\eps\kax{\eps a}^{r-1},
\ \text{for}\ |\beta|=1,
\end{equation}
and, if $\eps|a|\geq c_L$ then
\begin{equation}\label{eq:Vfar_2}
V(a)\geq c_V(\eps|a|)^r.
\end{equation}
To obtain the desired estimates on $a$ and $\mom$ we will use 
the fact that the soliton energy,
\begin{equation}
h(a,p):=\frac{1}{2}\big(p^2+V(a)\big),
\end{equation}
is essentially conserved.
We abbreviate $\alpha:=\{\ptr,\pbo,\alpha_{2d+1},\alpha_{2d+2}\}$.
The size of $\alpha$ is measured by $|\alpha|^2:=\sum_j |\alpha_j|^2$ 
and $\Anrm{\alpha}:=\sup_{s\leq t}|\alpha(s)|$.
We have the following:
\begin{proposition}\label{prop:apest2}
Let $V$ satisfy conditions \eqref{eq:Vup_2} and
\eqref{eq:Vfar_2}. Let $h_0:=h(\aZ,\mZ)$, and set
\begin{equation}
\label{eq:T11}
\tilde{T}_1:=\frac{C_{T_1}}{(\eps^2+\Anrm{\alpha})(1+\eps+h_0)}, 
\ \ C_{\tilde{T}_1}:=\frac{c_V}{2^{\max(2,r-1)/2}C_Vd},
\end{equation}
where the constants $C_V$ and $c_V$ are related to the growth rate of the
potential (see \eqref{eq:Vup} and \eqref{eq:Vfar}).
Then for times $t\leq \tilde{T}_1$:
\begin{equation}\label{eq:apest2}
|\mom|\leq C_{\tilde{p}}(\sqrt{h_0}+\Anrm{\alpha}t+\eps) \ \text{and}\ 
\eps|a|\leq C_{a},
\end{equation}
where $C_{a}$ and $C_{\tilde{p}}$ depend only on $c_L$, $c_V$,
$C_{\tilde{T}_1}$, $r$, $d$, $C_3$ and 
$C_E = \max(\eps,\epsE,\epsH)$.
$C_3$ is the constant in Corollary~\ref{cor:wo} and 
\end{proposition}
\begin{proof}
First we estimate $p$ in terms of $a$, using the almost conservation
of $h(a,p)$
\begin{equation}
\frac{d}{dt} h(a,p)= \frac{1}{2}\left(2p\cdot \left(\dot{p}+\nabla V(a)\right)+\nabla V(a)\cdot (\dot a-2p)\right).
\end{equation}
Now recall the definitions $\pbo:=-\dot p-\nabla V(a)$ and
$\ptr:=\dot a-2p$ together with the upper bound \eqref{eq:Vup_2} of
the potential $|\nabla V|\leq d^{1/2}C_V \eps \kax{\eps a}^{r-1}$
to obtain
\begin{equation}
|\dt h(a,p)|\leq  |\alpha||p| +\frac{1}{2}C_V d^{1/2}\eps |\alpha| \kax{\eps a}^{r-1}.
\end{equation}
Integration in time and simplification gives
\begin{equation}\label{eq:hbd}
h(a(t),p(t))\leq h_0 + t(\Anrm{\alpha})\left(\Anrm{p}+2^{-1}d^{1/2}C_V\eps \kax{\eps\Anrm{a}}^{r-1}\right).
\end{equation}
Recall that $h=2^{-1}(p^2+V(a))$ and that $V\geq 0$, thus 
$|p|^2 \leq 2h$.  
Solving the resulting quadratic inequality for $\Anrm{p}>0$ we find
that
\begin{equation}\label{eq:pbd}
\Anrm{p}\leq \sqrt{2 h_0} + 3t\Anrm{\alpha}+2^{-1}d^{1/2}C_V\eps\kax{\eps\Anrm{a}}^{r-1}.
\end{equation}
The Eqn.~\eqref{eq:hbd} also implies
\begin{equation}\label{eq:Vint}
\sup_{s\leq t}V(a(s))\leq 2h_0 + 2t\Anrm{\alpha}
\left(\Anrm{p}+2^{-1}d^{1/2}C_V\eps\kax{\eps\Anrm{a}}^{r-1}\right).
\end{equation}
As can be seen in \eqref{eq:pbd} we need to consider the possibility
of large $\eps|a|$.  Let $\eps|a|\geq c_L$, with $c_L$ as in
\eqref{eq:Vfar_2} then $V(a)\geq c_V(\eps|a|)^r$. Inserting this lower
bound and \eqref{eq:pbd} into \eqref{eq:Vint} we obtain
\begin{equation}
c_V(\eps\Anrm{a})^{r}\leq 2h_0  + 2t\Anrm{\alpha}\left(
\sqrt{2h_0} + 3t\Anrm{\alpha}+d^{1/2}C_V\eps\kax{\eps\Anrm{a}}^{r-1}
\right).
\end{equation}
Lemma~\ref{lem:maxmin} shows $\kax{\eps\Anrm{a}}^{r-1}\leq
2^{\max(0,r-3)/2}(1+(\eps\Anrm{a})^{r-1})$ for $r\geq 1$.
If the maximal time satisfies
the inequality $t\leq \tilde{T}_1$ (see~(\ref{eq:T11})),
then the above inequality implies
\begin{equation}\label{eq:tCa}
\eps\Anrm{a} \leq (\frac{2}{c_V}(C_4 + 2C_{\tilde{T}_1}+6C_{\tilde{T}_1}^2+
\frac{1}{2}c_V)^{1/r}=:\tilde{C}_a,
\end{equation}
where we have used that $h_0$ is bounded by the constant $C_E$.  Thus,
either $\eps|a|\leq c_L$ holds or, for the given time interval,
\eqref{eq:tCa} holds. In both cases $\eps|a|\leq C_a$, where
the constant only depends on $C_4=C_3C_E$, 
$C_{\tilde{T}_1}$, $c_V$, $c_L$ and $r$. 
We insert this upper bound on $\eps|a|$ into \eqref{eq:pbd} and
for times $t\leq \tilde{T}_1$ we find
\begin{equation}
\Anrm{p}\leq C_{\tilde{p}}(\sqrt{h_0}+\Anrm{\alpha}t+\eps), 
\end{equation}
where $C_{\tilde{p}}:=3+d^{1/2}C_VC_{\tilde{a}}^{r-1}$.
\end{proof}

Using the Corollary~\ref{cor:wo} we express the above proposition in 
terms of $\epsH$ rather than $h_0$. Recall the requirement on $\delta$ from
Theorem~\ref{thm:splitt}
\begin{corollary}\label{cor:apest}
  Let $V$ satisfy \eqref{eq:Vup}--\eqref{eq:Vfar} and let $\psi_0\in
  U_{\delta}\cap \Espace$. Furthermore, let $\psi_0$ satisfy the
  $\epsE$-energy bound \eqref{eq:ebnd} for $\sol_{\Par_0}$ with
  $\Par_0=\{a_0,p_0,\gamma_0,\freq_0\}$, and let $h(a_0,p_0)\leq
  \epsH$ (\ie \eqref{eq:hbound}). Let  
\begin{equation}
\label{eq:T22}
T_1:=\frac{C_{T_1}}{(\eps^2+\Anrm{\alpha})(1+\eps+\epsH+\eps)}, 
\ \ 
T_2:=\frac{\sqrt{\epsH}}{\Anrm{\alpha}+\eps^2},
\end{equation}
where
\begin{equation}
C_{T_1}:=\frac{C_{\tilde{T}_1}}{(1+C_3)(1+C_E^2)}.
\end{equation}
Then for times $t\leq \min(T_1,T_2)$:
\begin{equation}\label{eq2:apest}
|\mom|\leq C_p(\sqrt{\epsH}+\epsE+\eps) \ \text{and}\ 
\eps|a|\leq C_a,
\end{equation}
where $C_p$ depends on 
$C_E = \max(\eps,\epsE,\epsH)$,
$C_V$, $d$, $r$ and $C_a$. $C_3$ is defined in Corollary~\ref{cor:wo}
  and $C_a$ in Proposition~\ref{prop:apest2}. The constant $C_V$ is
  defined in \eqref{eq:Vup}.
\end{corollary}
\begin{proof}
  Under the assumptions of the corollary we have that
  Corollary~\ref{cor:wo} holds and hence
\begin{equation}
h(\aZ,\mZ)\leq C_3(\epsH+\epsE^2+\eps\epsE).
\end{equation}
We now modify the constants and estimates of Proposition~\ref{prop:apest2}
to take the upper bound of $h_0$ into account. 
The new, maximal time derived from $\tilde{T}_1$ becomes
$T_1 \leq \tilde{T}_1$.
For times shorter than this time, $t\leq T_1$, the bound on $\eps|a|$
remains the same.  Using this estimate for $\eps|a|$, we simplify the
$|p|$ estimate. Note first that $\sqrt{h_0}+\eps \leq
(\sqrt{\epsH}+\epsE+\eps)(1+2\sqrt{C_3})$, inserted into
\eqref{eq:apest2} gives
\begin{equation}
|p|\leq \frac{1}{2}C_{p}(\sqrt{\epsH}+\eps+\epsE+|\alpha|t),
\end{equation}
where $C_p$ depends on $C_3$, $C_E$, $C_a$ and $d$ and $r$.
With the choice of time interval $T_2$ such that $t\leq T_2$, where
$T_2$ is given in~(\ref{eq:T22}),
we obtain $|p|\leq C_p(\sqrt{\epsH}+\epsE+\eps)$.
\end{proof}

\section{Lyapunov functional}
\label{sec:4}

In this section we define the Lyapunov functional and calculate its time
derivative in the moving frame. Recall the definition of $\Ew(\psi)$ in
\eqref{eq:Ew} together with decomposition \eqref{eq:splitt}:
$\psi= \TB(\solw+w)$, with  $w\bot J \TM$.  Define the
Lyapunov functional, $\Lambda$, as
\begin{equation}\label{eq:L}
\Lambda := \Ew(\solw+w) + \frac{1}{2}\dotp{\VR (\solw+w)}{\solw+w} 
- \Ew(\solw) - \frac{1}{2}\dotp{\VR\solw}{\solw}.
\end{equation}

Here we show that the Lyapunov functional $\Lambda$ is an almost
conserved quantity. We begin by computing its time derivative.  Let
$\pbo:=-\dot\mom-\nabla V(a)$ and $\ptr:=\dot a-2\mom$ (boost and
translation coefficients).  We have the following proposition
\begin{proposition}\label{prop:dL}
Given a solution $\psi\in \Espace\cap U_\delta$ to \eqref{eq:NLS}, define 
$\solw$ and $w$ as above. Then
\begin{equation}
\frac{d}{dt} \Lambda = \mom \cdot
\dotp{\nabla_a \VR w}{w} - \ptr\cdot \mathrm{D}^2V(a)\cdot
\dotp{xw}{w} + R ,
\end{equation}
where 
\begin{equation}
\begin{split}
R &: =\pbo\cdot\dotp{\iu w}{\nabla w} + 2\mom \cdot \dotp{\nabla_a
\VR\solw}{w}-\frac{1}{2}\ptr \cdot \dotp{\nabla_a\VR\solw}{\solw} \\ 
&+ \frac{\dot\freq}{2}\nrm{w}^2  -
\dot\freq\dotp{\VR\solw}{\partial_\freq\solw}.
\end{split}
\end{equation}
\end{proposition}
Before proceeding to the proof, we recall the definition of the moving
frame solution $u$ defined by
\begin{equation}\label{eq:u}
u(x,t):= \lexp{-\iu \mom \cdot x - \iu \gamma}\psi(x+a,t).
\end{equation}
Here $a$, $\mom$ and $\gamma$ depend on time, in a way
determined by the splitting of Section~\ref{sec:3},
and the function $\psi$ is a solution of the nonlinear
Schr\"odinger equation \eqref{eq:NLS}. In the moving frame the
Lyapunov functional $\Lambda$ takes the form
\begin{equation}\label{eq:Lt}
\Lambda = \Ew(u) + \frac{1}{2}\dotp{\VR u}{u} 
- \Ew(\solw) - \frac{1}{2}\dotp{\VR\solw}{\solw}.
\end{equation}

We begin with some auxiliary lemmas.
\begin{lemma}\label{lem:Ehr}
Let $\psi\in \Espace$ be a solution to \eqref{eq:NLS}. Then
\begin{equation}\label{eq:Ehr}
\partial_t \dotp{\psi}{-\iu \nabla \psi} = - \dotp{(\nabla V)\psi}{\psi}
\ \text{and}\ \
\partial_t\dotp{x \psi }{\psi}= 2\dotp{\psi}{-\iu \nabla \psi}.
\end{equation}
\end{lemma}
\begin{proof}
The first part of this lemma was proved in \cite{FGJS-I}. 
To prove the second part we use the equation
\begin{equation}
\partial_t(x_k|\psi|^2) = \iu \nabla \cdot (x_k
\bar{\psi}\nabla \psi - x_k \psi\nabla \bar{\psi}) -
\iu(\bar{\psi}\partial_k \psi - \psi \partial_k \bar{\psi}),
\end{equation}
understood in a weak sense, which follows from the nonlinear
Schr\"odinger equation \eqref{eq:NLS}. Formally, integrating this
equation and using that the divergence term vanishes gives the second
equation in \eqref{eq:Ehr}. To do this rigorously, let $\chi$ be a
$\C{1}$ function such that $|\nabla \chi(x)|\leq C$ and
\begin{equation}
\chi(x):=\left\{\begin{array}{ll} 1 & |x|\leq 1, \\ 0 & |x|>2,
\end{array}\right.
\end{equation}
and let $\chi_R(x):=\chi(\frac{x}{R})$.
Abbreviate $j_k:=(x_k \bar{\psi}\nabla \psi - x_k \psi\nabla
\bar{\psi})$ and let $R>1$. We multiply the divergence term by
$\chi_R$. Integration by parts gives
\begin{equation}
\left|\int (\nabla\cdot j_k) \chi_R\diff^d x\right|=\left|\int j_k\cdot \nabla
\chi_R(x) \diff^d x\right| \leq \frac{C}{R} \int |j_k| \diff^d x.
\end{equation}
We note that $j_k\in \Lp{1}$ for all $k$, and is independent of $R$, thus as
$R\rightarrow \infty$, this term vanishes. The remaining terms give in
the limit $R\rightarrow \infty$ the second equation in \eqref{eq:Ehr}.
\end{proof}
\begin{lemma}\label{lem:dEu}
Let $\psi\in\Espace$ be a solution to \eqref{eq:NLS}, and let $u$ be defined
as above. Then
\begin{equation}
\begin{split}
\frac{d}{dt} \big(\Ew(u)+\frac{1}{2}\dotp{\VR u}{u}\big) &= 
\mom \cdot \dotp{\nabla_a \VR u}{u} 
- \frac{1}{2} \ptr  \cdot \mathrm{D}^2 V(a) \cdot  \dotp{x u}{u} \\ 
&+ \frac{1}{2}\dot\freq \nrm{u}^2 + \pbo\cdot \dotp{\iu u}{\nabla u},
\end{split}
\end{equation}
where $\ptr := \dot a - 2 \mom$ and $\pbo =-\dot\mom -\nabla V(a)$.
\end{lemma}
\begin{proof}
The functional $\Ew(u)+\frac{1}{2}\dotp{\VR u}{u}$, is related to the
Hamiltonian functional by
\begin{equation}\label{eq:Ku}
\begin{split}
\Ew(u) + \frac{1}{2}\dotp{\VR u}{u} &= 
\HV(\psi) + \frac{1}{2}(\mom^2+\freq)\nrm{\psi}^2 - 
\mom \cdot \dotp{\iu \psi}{\nabla \psi} \\ &- 
\frac{1}{2}\int (V(a)+\nabla V(a)\cdot (x-a))|\psi|^2 \diff^d x ,
\end{split}
\end{equation}
which is obtained by substituting \eqref{eq:u} into
$\Ew(u)+\frac{1}{2}\dotp{\VR u}{u}$.
Using the facts that the mass $\nrm{\psi}^2$ and 
Hamiltonian $\HV(\psi)$ are time independent,
together with the Ehrenfest relations, Lemma~\ref{lem:Ehr}, we
obtain
\begin{equation}\nonumber
\begin{split}
\frac{d}{dt} \big(\Ew(u)+\frac{1}{2}\dotp{\VR u}{u}\big) &=
(\frac{\dot{\freq}}{2} + \mom\cdot\dot\mom)\nrm{\psi}^2 - \dot
\mom\cdot\dotp{\iu \psi}{\nabla \psi} + \mom\cdot \dotp{(\nabla
V)\psi}{\psi} \\ &- \frac{\dot a}{2}\cdot \mathrm{D}^2V(a)\cdot \int (x-a)|\psi|^2 \diff^d x -
\nabla V(a)\cdot \dotp{\iu \psi}{\nabla \psi}.
\end{split}
\end{equation}
Collecting $\mom\cdot\dot\mom$ and $\mom\cdot \nabla V$ together, and 
combining $\dot\mom$ and $\nabla V(a)$ gives
\begin{multline}\label{eq:dt1}
\frac{d}{dt} \big(\Ew(u)+\frac{1}{2}\dotp{\VR u}{u}\big) =
\frac{\dot\freq}{2}\nrm{\psi}^2 + 
\mom\cdot\dotp{(\dot\mom+ \nabla V)\psi}{\psi} \\
- (\dot\mom + \nabla V(a))\cdot\dotp{\iu\psi}{\nabla
\psi} - \frac{1}{2}\dot a\cdot 
\mathrm{D}^2 V(a)\cdot \int (x-a) |\psi|^2
\diff^d x.
\end{multline}
From the definition of $u$, \eqref{eq:u}, the following relations hold
\begin{eqnarray}\label{eq:urel}
&\nrm{\psi} = \nrm{u},\qquad \dotp{\iu \psi}{\nabla \psi} = p\nrm{u}^2+\dotp{\iu
u}{\nabla u},&  \\ 
&\dotp{(\nabla V)\psi}{\psi} = \dotp{(\nabla V_a) u}{u}, \qquad 
\dotp{(x-a)\psi}{\psi} = \dotp{xu}{u}.& \label{eq:ures}
\end{eqnarray}
Substitution of \eqref{eq:urel}--\eqref{eq:ures} into \eqref{eq:dt1}
gives, after cancellation of the $\mom\cdot \dot\mom$ terms,
\begin{multline}
\frac{d}{dt} \big(\Ew(u)+\frac{1}{2}\dotp{\VR u}{u}\big) =
\frac{\dot\freq}{2}\nrm{u}^2 + \mom\cdot\dotp{(\nabla V_a-\nabla
V(a))u}{u} \\ - (\dot\mom + \nabla V(a))\cdot \dotp{\iu u}{\nabla u} -
\frac{1}{2}\dot a\cdot \mathrm{D}^2 V(a)\cdot \int x |u|^2 \diff^d x.
\end{multline}
The last remaining step is to rewrite the second last term as $\dot a
-2\mom+2\mom$ and combine its $\mom$ term with the difference of the
potentials, recalling the definition of $\VR$, to obtain
\begin{multline}
\frac{d}{dt} \big(\Ew(u)+\frac{1}{2}\dotp{\VR u}{u}\big) =
\frac{\dot\freq}{2}\nrm{u}^2 + \mom\cdot\dotp{(\nabla_a \VR)u}{u} 
\\ -
(\dot\mom + \nabla V(a))\cdot \dotp{\iu u}{\nabla u} +
\frac{1}{2}(2\mom-\dot a)\cdot \mathrm{D}^2 V(a)\cdot \int x |u|^2
\diff^d x.
\end{multline}
Identification of the boost coefficient $\pbo:=-\dot \mom - \nabla
V(a)$ and the translation coefficient $\ptr:=\dot a-2\mom$ gives
the lemma.
\end{proof}

The time derivative of the second part of
the Lyapunov functional \eqref{eq:Lt} is computed in the next lemma.
\begin{lemma}\label{lem:dEsol}
Let $\solw$ be the solution of \eqref{eq:sol}, and let $\freq$ depend on $t$.
Then
\begin{multline}
\frac{d}{dt} \big(\Ew(\solw) + \frac{1}{2}\dotp{\VR\solw}{\solw}\big) = \\
\frac{\dot\freq}{2}\nrm{\solw}^2 + 
(\mom + \frac{1}{2}\ptr)\cdot \dotp{\nabla_a \VR\solw}{\solw} + 
\dot\freq \dotp{\VR \solw}{\partial_\freq \solw},
\end{multline}
where $\ptr:=\dot a -2\mom$.
\end{lemma}
\begin{proof}
The result follows directly, upon recalling that $\DE(\solw)=0$ and
$\frac{1}{2}\ptr + \mom=\frac{\dot a}{2}$.
\end{proof}

To proceed to the proof of Proposition~\ref{prop:dL}, we restate our
condition for unique decomposition of the solution to the
nonlinear Schr\"odinger equation, $\psi\in U_\delta \cap \Espace$, in
terms of $u$:
\begin{equation}\label{eq:usplitt}
u = \solw+w\quad\text{and}\quad w\bot J \TM.
\end{equation}
Given Lemma~\ref{lem:dEu} and Lemma~\ref{lem:dEsol}, Proposition~\ref{prop:dL}
follows directly.
\begin{proof}[Proof of Proposition~\ref{prop:dL}]
Lemma~\ref{lem:dEu} states
\begin{equation}
\begin{split}
\frac{d}{dt} \big(\Ew(u)+\frac{1}{2}\dotp{\VR u}{u}\big) &= 
\mom \cdot \dotp{\nabla_a \VR u}{u}
- \frac{1}{2} \ptr  \cdot \mathrm{D}^2 V(a) \cdot  \dotp{x u}{u} \\ 
&+ \frac{1}{2}\dot\freq \nrm{u}^2 + \pbo\cdot \dotp{\iu u}{\nabla u}.
\end{split}
\end{equation}
Insert $u=\solw+w$ above, and use $w\bot \{\solw$, $\iu
\nabla \solw$, $x\solw\}$. Recall that $\solw$ is a real valued symmetric
function, hence $\dotp{x\solw}{\solw}=0$ as well as $\dotp{\iu
\solw}{\nabla \solw}=0$. We obtain
\begin{multline}
\frac{d}{dt} \big(\Ew(u)+\frac{1}{2}\dotp{\VR u}{u}\big) = \\
\mom \cdot \big(\dotp{\nabla_a
\VR w}{w} + 2\dotp{\nabla_a \VR \solw}{w} + \dotp{\nabla_a \VR
\solw}{\solw}\big)\\ - \frac{1}{2} \ptr \cdot \mathrm{D}^2 V(a)
\cdot \dotp{x w}{w} + \frac{1}{2}\dot\freq (\nrm{w}^2+\nrm{\solw}^2) +
\pbo\cdot \dotp{\iu w}{\nabla w}
\end{multline}
Subtracting the result of Lemma~\ref{lem:dEsol} we find
\begin{multline}
\frac{d}{dt} \Lambda =
\mom \cdot \dotp{(\nabla_a \VR)w}{w} 
- \frac{1}{2}\ptr\cdot \mathrm{D}^2 V(a)\cdot \dotp{xw}{w} \\
+ \pbo \cdot \dotp{\iu w}{\nabla w} 
+ 2\mom \cdot \dotp{\nabla_a \VR \solw}{w} - \frac{1}{2}\ptr \cdot \dotp{\nabla_a \VR \solw}{\solw} + \frac{\dot\freq}{2}\nrm{w}^2 \\
- \dot\freq \dotp{\VR \solw}{\partial_\freq\sol} .
\end{multline}
Note that the terms on the second and third line are at least fourth
order in the small parameters.  The last two lines is the definition
of $R$ in the proposition.
\end{proof}

\section{Upper bound on $\Lambda$}
\label{sec:6}

This section we estimate $\Lambda$ from above using
Corollary~\ref{cor:apest} in Proposition~\ref{prop:dL}.  Taylor
expansion of $\Ew\big(\sol(t)+w(x,t)\big)$ around $\sol$ at
$t=0$, gives 
\begin{equation}\label{eq:Eb}
|\En_{\freq(t)}(u(x,t)) - \Ew(\sol_{\freq(t)}(x))|_{t=0}\leq C \nrmHo{\wZ}^2.
\end{equation}
The remaining terms in the
Lyapunov functional are estimated using the inequality $\Hess V(x)\leq
C\eps^2|x|^2\kex^{r-2}$ together with Taylor's formula and
Lemma~\ref{lem:VRup2}.  Furthermore, we use from Corollary~\ref{cor:apest}. that $|\eps \aZ|\leq C$. We obtain for a $\theta\in[0,1]$
\begin{multline}\label{eq:Rb}
\left|\dotp{\VR u}{u)}-\dotp{\VR\sol}{\sol}\right|_{t=0} =
\left|\dotp{\VR w}{w)}+2\dotp{\VR\sol}{w}\right|_{t=0} \\=
\eps^2|\dotp{x\cdot \Hess V(x\theta+\aZ)\cdot x}{2\sol_{\freq_0}
\RE(w_0)} +|\dotp{\VR \wZ}{\wZ}| \\ 
\leq C(\eps^2\nrm{\wZ} + \nrm{\wZ}^2+\Bnrm{\eps x \kex^{(r-2)/2}\wZ}^2).
\end{multline}

We now use Corollary~\ref{cor:wo} and Lemma~\ref{lem:eqv} in \eqref{eq:Rb} 
and \eqref{eq:Eb} to obtain
\begin{equation}\label{eq:L2}
\left|\dotp{\VR u}{u)}-\dotp{\VR\sol}{\sol}\right|_{t=0}
\leq C(\eps^2 \epsE + \epsE^2)
\end{equation}
and 
\begin{equation}
|\En_{\freq(t)}(u(x,t)) -
\Ew(\sol_{\freq(t)}(x))|_{t=0}\leq C\epsE^2.
\end{equation}
Thus, finally
\begin{equation}\label{eq:L0}
|\Lambda|_{t=0}\leq C (\epsE^2 + \eps^2\epsE).
\end{equation}

\begin{proposition}\label{prop:better}
  Let $\psi\in U_{\delta}\cap \Espace$, and let $\Lambda$, $w$ and
  $\alpha$ be defined as above, and $\delta$ as defined in
  Theorem~\ref{thm:splitt}. Then
\begin{multline}
|\frac{d}{dt}\Lambda|\leq C\Big((\eps+\epsE
+\sqrt{\epsH})\eps\nrm{\eps x\kex^{(r-2)/2}w}^2+|\alpha|\eps\Bnrm{(\eps|x|)^{1/2}w}^2
 \\ + \big((\eps+\epsE+\sqrt{\epsH})\eps^2+ |\alpha|\big)(\nrmHo{w}^2+
\eps^2)\Big),
\end{multline}
for times $0\leq t\leq \min(T_1,T_2)$, where $T_1$ and $T_2$ are defined in 
Corollary~\ref{cor:apest}.
\end{proposition}
\begin{proof}
Proposition~\ref{prop:dL} implies
\begin{multline}\label{R}
|\frac{d}{dt}\Lambda| \leq C \big( |p| |\dotp{\nabla_a \VR w}{w}| +
|\ptr||\Hess V(a)| |\dotp{xw}{w}| \\ + |\pbo|\nrm{w}\nrm{\nabla w} +
|p|\eps^3 \nrm{w} + |\ptr|\eps^3 + |\dot\freq|\nrm{w}^2 +
|\dot\freq|\eps^2\big).
\end{multline}
An alternative form of Eqn.~\eqref{R} is
\begin{multline}\label{eq:alt}
|\frac{d}{dt} \Lambda| \leq C \big( |p| |\dotp{\nabla_a \VR w}{w}| 
+ |\alpha||\Hess V(a)| |\dotp{xw}{w}| \\ + 
 (|\mom|\eps^2+ |\alpha|)(\nrmHo{w}^2+ \eps^2)\big),
\end{multline}
where we have used $\eps<C$ and $|\alpha_j|\leq |\alpha|$, $\forall j$.

Using Corollary~\ref{cor:RVup} we estimate the $\VR$ terms to obtain
\begin{multline}
  |\frac{d}{dt} \Lambda| \leq C \big(
 |p|\eps
  \nrm{\eps|x|\kex^{(r-2)/2}w}^2  +
  |\alpha|\eps\kax{\eps\Anrm{a}}^{r-2}|\dotp{\eps xw}{w}| \\ +
  (|\mom|\eps^2+ |\alpha|)(\nrmHo{w}^2+ \eps^2)\big).
\end{multline}
The proposition now follows upon using $\eps|a|\leq C_a$ and $|p|\leq
C(\eps+\epsE+\sqrt{\epsH})$ for $t\leq \min(T_1,T_2)$ from
Corollary~\ref{cor:apest} and the inequality:
\begin{equation}
\dotp{\eps xw}{w}\leq \nrm{(\eps|x|)^{1/2}w}^2.
\end{equation}
\end{proof}

Equation~\eqref{eq:L0} and Proposition~\ref{prop:better} yield an
upper bound on $\Lambda$:
\begin{equation}\label{eq:Lup}
|\Lambda| \leq C\epsE^2 + C\eps^2\epsE + 
t\sup_{s\leq t} |\frac{d}{dt} \Lambda|.
\end{equation}

\section{Lower bound on $\Lambda$}
\label{sec:7}

In this section we estimate the Lyapunov-functional $\Lambda$ from
below.  Recall the definition \eqref{eq:L} of $\Lambda$:
\begin{equation}
\Lambda:=\Ew(\sol+w)-\Ew(\sol) + 
\frac{1}{2}\dotp{\VR(\sol+w)}{\sol+w} - 
\frac{1}{2}\dotp{\VR\sol}{\sol}.
\end{equation}
We have the following result.
\begin{proposition}\label{prop:Llow}
Let $\Lambda$ and $w$ be defined as above.
Then for a positive constant $C$,
\begin{equation}\label{eq:Lcoer}
\Lambda\geq \frac{1}{2}\rho_2 \nrmHo{w}^2 + C_0\rho_1\nrm{\eps|x|
\kex^{(r-2)/2}w}^2 - C\nrmHo{w}^3-C\eps^2\nrm{w}.
\end{equation}
where $r$ and $\rho_1>0$ are defined in \eqref{eq:Vup}, $C_0$ is the
positive constant defined in Lemma~\ref{lem:RVbd} and $\rho_2>0$ is a
positive number. The constant $C_0$ depends on the constant $C_a$ defined in
Corollary~\ref{cor:apest} bounding the size of $\eps|a|$.
\end{proposition}
\begin{proof}
By Taylor expansion we have
\begin{equation}
\Ew(\sol+w)-\Ew(\sol) = \frac{1}{2}\dotp{\LL w}{w} + \RIII{w},
\end{equation}
where $\LL:=(\Hess \Ew)(\sol)$ and
by Condition~\ref{con:A}, $|\RIII{w}|\leq C\nrmHo{w}^3$.  The
coercivity of $\LL$ for $w\bot J\TM$ is proved in Proposition D.1
of \cite{FGJS-I} under 
Conditions~\ref{con:GWP}--\ref{con:F} on the
nonlinearity (in Section~\ref{sec:ass}). Thus
\begin{equation}\label{eq:bl1}
\dotp{\LL w}{w}\geq \rho_2 \nrmHo{w}^2 \ \text{for}\ w\bot J\TM.
\end{equation}

The remaining terms of $\Lambda$ can be rewritten as
\begin{equation}
\dotp{\VR(\sol+w)}{\sol+w}-\dotp{\VR\sol}{\sol} =
\dotp{\VR w}{w}+2\dotp{\VR\sol}{w}.
\end{equation}
In Lemma~\ref{lem:RVbd} we show that 
\begin{equation}\label{eq:bl2}
\VR \geq C_0\rho_1(\eps|x|)^2\kex^{r-2} \ \text{for}\ r\geq 1.
\end{equation}
Using Lemma~\ref{lem:RVbd}, \eqref{eq:bl1},  \eqref{eq:bl2} and the fact that
$\dotp{\VR\sol}{w}\leq C\eps^2\nrm{w}$ we obtain the lower bound on $\Lambda$.
\end{proof}

\section{Proof of Theorem~\ref{thm:main}}
\label{sec:end}

The upper bound \eqref{eq:Lup} together with the bound from below in
Proposition~\ref{prop:Llow} yield the inequality
\begin{multline}\label{eq:ml}
\frac{1}{2}\rho_2 \nrmHo{w}^2 + C_0\rho_1
\Bnrm{\eps x\kex^{(r-2)/2}w}^2 -
C\nrmHo{w}^3-C\eps^2\nrm{w}\leq  C\epsE^2 + C\eps^2\epsE \\ + tC \sup_{s\leq
t}\Big(
(\epsT+\sqrt{\epsH})\eps\nrm{\eps x\kex^{(r-2)/2}w}^2+
|\alpha| \eps\nrm{(\eps |x|)^{1/2}w}^2 \\+ ((\epsT+\sqrt{\epsH})\eps^2+
|\alpha|)(\nrmHo{w}^2+ \eps^2)\Big),
\end{multline}
for $0\leq t\leq \min(T_1,T_2)$, where $T_1$ and $T_2$ are defined in
Corollary~\ref{cor:apest} and $\epsT:=\eps+\epsE$.  
The right-hand side is independent
of the operator $t\mapsto s$, $\sup_{s\leq t}$ in the given time interval, 
we can therefore apply this to
both sides of \eqref{eq:ml}. To simplify, let
\begin{equation}
\rho:=\min(\frac{\rho_2}{8},\frac{C_0\rho_1}{3}).
\end{equation} 
We absorb higher order terms into lower order ones. Furthermore, we assume
\begin{equation}
t \leq \min(T_1,T_2,T_3), \ \text{where}\ 
T_3:=\frac{\rho}{C(\Anrm{\alpha}+\eps(\epsT+\sqrt{\epsH}))
(1+\eps)},
\end{equation}
in agreement with Corollary~\ref{cor:apest}. Both $\rho$ and $C$ above depend on $I$, clarifying the need for $\epsT\ll C(I)$. Note that
\begin{equation}
T_3C(\epsT+\sqrt{\epsH})\eps\leq \rho, \ T_3C\Anrm{\alpha}\eps\leq \rho, \
\text{and}\ T_3C((\epsT+\sqrt{\epsH})\eps^2+\Anrm{\alpha}\leq 2\rho.
\end{equation}
We obtain
\begin{multline}\label{eq:fint}
  \rho \sup_{s\leq t} \left(4\nrmHo{w}^2 + 3\Bnrm{\eps
      x\kex^{(r-2)/2}w}^2 \right)\\  \leq C \big( \sup_{s\leq t} (
    \nrmHo{w}^3+\eps^2\nrm{w})+ \epsE^2 + \eps^2\epsE\big) \\ + 
    \rho\sup_{s\leq t} \Big( \Bnrm{\eps x\kex^{(r-2)/2}w}^2+
    \Bnrm{|\eps x|^{1/2}w}^2+2\eps^2 + 2\nrmHo{w}^2\Big).
\end{multline}
Note that $g(y):=|y|-y^2\kax{y}^{-1}\leq 2^{-1}$, $y\in\mathbb{R}$.
Indeed $g(-y)=g(y)$ and $g$ is continuously differentiable on
$(0,\infty)$, $g(y)\geq 0$ since $|y|\geq y^2\kax{y}^{-1}$ with
$g(0)=g(\infty)=0$. The function $g(y)$ has one critical point on
$(0,\infty)$ at $y=(2^{-1}(\sqrt{5}-1))^{-1/2}$ with value $\max
g=(3-\sqrt{5})(2(\sqrt{5}-1))^{-1/2}\leq 2^{-1}$. This proves the claim. 
We now use this intermediate function $g(x)$ to estimate the term 
above with $|x|^{1/2}$. We have
\begin{equation}\label{eq:gex}
\eps |x|-(\eps |x|)^2\kex^{r-2}\leq g(\eps|x|)\leq \frac{1}{2}.
\end{equation}
We also have the inequalities
\begin{equation}
C\nrmHo{w}^3\leq 
\rho^{-1}C^2\nrmHo{w}^4+4^{-1}\rho\nrmHo{w}^2, \ \
C\eps^2\nrmHo{w}\leq C^2\rho^{-1}\eps^4+4^{-1}\rho\nrmHo{w}^2.
\end{equation}
Thus we have $3\rho\nrmHo{w}^2$ on the right-hand side and
$2\rho$ of terms containing $\kex$. Moving those to the 
left-hand side of \eqref{eq:fint} using the above inequalities 
and simplifying we obtain
\begin{equation}\label{eq:fint2}
  \sup_{s\leq t} \left(\nrmHo{w}^2 + \Bnrm{\eps
      x\kex^{(r-2)/2}w}^2 \right) \leq C'\epsT^2 + 
    C^2\rho^{-2}(\sup_{s\leq t}\nrmHo{w}^4).
\end{equation}
Abbreviate
$\lite:=C'\epsT^2$.
Let 
\begin{equation}
X:=\sup_{s\leq t}\left(\nrmHo{w}^2 + \Bnrm{\eps|x|\kex{\eps
    x}^{(r-2)/2}w}^2\right).
\end{equation} 
Equation~\eqref{eq:fint2} implies
\begin{equation}
X \leq  C^2\rho^{-2}X^2+ \lite.
\end{equation}
Solving this inequality, we find
\begin{equation}\label{eq:eta}
X\leq 2
\lite,\ \text{provided}\ \lite\leq \frac{\rho^2}{4C^2}.
\end{equation}
The definition of $X$ and $\lite$ implies
\begin{equation}\label{eq:1012}
\nrmHo{w}\leq c'\epsT, \ \ \text{and}\
\Bnrm{\eps x\kex{\eps x}^{(r-2)/2}w}\leq c' \epsT.
\end{equation}
Lemma~\ref{lem:eqv} allow us to rewrite \eqref{eq:1012} as
$\Enrm{w}\leq c'\epsT$. Inserting \eqref{eq:1012} into the expressions
for our modulation parameters, the estimate of the $\alpha_j$-terms in
\eqref{eq:gam}--\eqref{eq:bo} gives us $|\alpha|\leq c\epsT^2$ and
time interval $t\leq T'$, where
\begin{equation}
T' := c\min(\epsT^{-2},\frac{\sqrt{\epsH}}{\epsT^2},
\frac{1}{\epsT^2+\eps\sqrt{\epsH}})
\end{equation}
Using $\epsH\geq K\eps$ (that is, $\epsH$ is not 
an order of magnitude smaller then $\eps$), we can shorten
the time-interval to have an upper limit of
\begin{equation}
T'':= C (\epsT^2+\eps\sqrt{\epsH})^{-1}.
\end{equation}
We now choose $\epsT$ such that \eqref{eq:eta} holds and $c'\epsT \leq
\frac{1}{2}\delta$, where $\delta$ is defined in
Theorem~\ref{thm:splitt}. Then there
is a maximum $T_0$ such that the solution $\psi$ of \eqref{eq:NLS} is
in $U_{\delta}$ for $t\leq T_0$.  Thus the decomposition
\eqref{eq:splitt} is valid and the above upper bounds for $\nrmHo{w}$
and $\alpha$ are valid for $t\leq
\min(T_0,C(\epsT^2+\eps\sqrt{\epsH})^{-1})$. Thus there exists a
constant $C_T$ such that $0<C_T\leq C$, such that for $t\leq
C_T(\epsT^2+\eps\sqrt{\epsH})^{-1}$ the theorem holds.  This concludes
the proof of Theorem~\ref{thm:main}.\hfill\qed

\appendix

\section{Lower bound on $\VR$}
\label{app:RVbd}

In this appendix we estimate $\VR$ from below. Recall that 
\begin{equation}
\VR(x):=V(x+a)-V(a)-\nabla V(a)\cdot x
\end{equation}
and 
\begin{equation}
\Hess V\geq \rho_1 \eps^2 \langle \eps x \rangle^{(r-2)/2},
\end{equation}
where $\rho_1$ is a positive constant.  We have
the following result:
\begin{lemma}\label{lem:RVbd}
Let $a,x\in\mathbb{R}^d$ and $0<\eps\in\mathbb{R}$. Then (i) if $r\geq 2$
or (ii) if $r\leq 2$ and $\eps|a|\leq C_a$:
\begin{equation}
\VR(x)\geq C_0\rho_1 \eps^2|x|^2\kex^{r-2},
\end{equation}
where
\begin{equation}
C_0:=\left\{\begin{array}{ll} 
\frac{1}{2^{r-2+\max(0,\frac{r-4}{2})}r(r-1)} & \text{in case (i)} \\ 
\frac{1}{2\left(2(1+C_a^2)\right)^{(2-r)/2}} & \text{in case (ii)}.
\end{array}\right.
\end{equation}
\end{lemma}
\begin{proof}
  Consider the case $x=0$, from the definition of $\VR$ it follows that
  $\VR(0)=0$, thus the estimate holds. Let $x\neq 0$, the function
  $\VR(x)$ is the Taylor expansion of $V(x+a)$ around $a$ to
  first order. The Taylor series remainder
\begin{equation}
  \int_0^1 (1-\theta)x\cdot \Hess V(a+\theta x)\cdot x\diff \theta,
\end{equation}
integrated by parts, gives the identity
\begin{multline}
  \int_0^1 (1-\theta)x\cdot \Hess V(a+\theta x)\cdot x \diff \theta
  = \\ \left.(1-\theta)x\cdot \nabla V(a+\theta x)\right|_0^1 +
  \int_0^1 \nabla V(a+\theta x)\cdot x\diff \theta=\VR(x,t).
\end{multline}
Inserting the lower bound of the $\Hess V$ into the Taylor remainder
gives the inequality
\begin{equation}\label{eq:LI}
\VR(x)\geq \rho_1\eps^2|x|^2\int_0^1 (1-\theta)
(1+\eps^2|a+x \theta|^2)^{(r-2)/2}\diff \theta = \rho_1 \eps^2 |x|^2 I.
\end{equation}

To estimate $I$, we first consider case (a), with $r\geq 2$. 
The integrand of $I$ is estimated by the following lemma.
\begin{lemma}\label{lem:c1}
Let $y\geq 0$ and $\be\geq 0$ then
\begin{equation}
\frac{1}{2^{\max(0,\frac{2-\be}{2})}}\leq \frac{(1+y^2)^{\be/2}}{1+|y|^\be}\leq 2^{\max(0,\frac{\be-2}{2})}.
\end{equation}
\end{lemma}
This lemma is proved at the end of this appendix. 
For $x\neq 0$ denote $\hat{x}=x/|x|$,
$\ap=a\cdot \hat{x}$, and $\ab=a-\ap\hat{x}$, and abbreviate $\be:=r-2$.
Lemma \ref{lem:c1} then implies
\begin{align}
\left(1+\eps^2|a+x \theta|^2\right)^{\be/2}&  
=\left(1+\eps^2 |\ab|^2\right)^{\be/2}
\left(1+\frac{\eps^2}{1+\eps^2|\ab|^2}
\left(\ap+|x|\theta\right)^2\right)^{\be/2} \nonumber \\
&\geq \frac{1}{2^{\max(0,\frac{2-\be}{2})}}\left(\left(1+\eps^2|\ab|^2\right)^{\be/2}
+\eps^{\be}\left|\ap+|x|\theta\right|^{\be}\right) \nonumber \\ &\geq
\frac{1}{2}+\frac{1}{2}\eps^{\be}\left|\ap+|x|\theta\right|^{\be},
\end{align}
where we used $2^{\max(0,\frac{2-\be}{2})}\leq 2$.  Thus
\begin{equation}\label{eq:I}
I\geq \frac{1}{4} + \frac{1}{2}\eps^{r-2}\int_0^1 (1-\theta)\big|\ap+|x|\theta\big|^{r-2}\diff \theta
=\frac{1}{4}+\frac{1}{2}\eps^{r-2}I_2.
\end{equation}
The integral $I_2$ evaluates to
\begin{equation}\label{eq:Iest}
I_2 =\frac{1}{r(r-1)|x|^2}\left(
\left|\ap+|x|\right|^r-|\ap|^r-r|x|\sign(\ap)|\ap|^{r-1}\right),
\end{equation}
which is the remainder of a Taylor series of $\big|\ap+|x|\big|^{r}$ around
$\ap$, as expected. At $\ap=0$ we have 
\begin{equation}\label{eq:I20}
I_2\geq \frac{|x|^{r-2}}{r(r-1)}.
\end{equation}
To estimate $I_2$ for $\ap\neq 0$, we use the following lemma
\begin{lemma}\label{lem:c3}
Let $y\in \mathbb{R}$ and $r\geq 2$. Then
\begin{equation}
\left|1+y\right|^r - 1 - r y\geq 
\frac{1}{2^{r-2}}|y|^r.
\end{equation}
\end{lemma}
The lemma is proved at the end of this appendix.  Since $\ap\neq 0$ we
can pull it out of $I_2$, and use $y:=|x|/\ap$ in the Lemma to
obtain
\begin{equation}\label{eq:I21}
I_2 \geq \frac{|x|^{r-2}}{2^{r-2}r(r-1)}.
\end{equation}
Note that the above result is a lower limit than we obtained in
\eqref{eq:I20}, so we can use \eqref{eq:I21} for all $\ap$. Now,
inserting this inequality into \eqref{eq:I} and the result into \eqref{eq:LI}
to obtain
\begin{equation}
\VR(x)\geq \rho_1 \eps^2 |x|^2 I\geq
\rho_1\left(\frac{\eps^2|x|^2}{4}+\frac{\eps^r|x|^r}{2^{r-1}r(r-1)}\right).
\end{equation}
Once again using Lemma \ref{lem:c1} gives
\begin{equation}
\VR(x)\geq C_1 \rho_1 \eps^2 |x|^2 \kex^{r-2},
\end{equation}
where 
\begin{equation}
C_1:=\frac{1}{2^{\max(0,\frac{r-4}{2})}2^{r-2}r(r-1)}, \ r\geq 2.
\end{equation}
This concludes part (i) of Lemma \ref{lem:RVbd}, except for the proofs
of Lemma \ref{lem:c1} and \ref{lem:c3} which is done below.

Now, we estimate the integral $I$ for the case (ii), with $r\leq 2$
and $\eps|a|\leq C_a$.  Introduce the change of variables
$p=(1-\theta)^2$. The integral takes the form
\begin{equation}
I=\frac{1}{2}\int_0^1 \frac{1}{g(p)^{(2-r)/2}}\diff p, \ 
\text{where}\ g(p):=1+\eps^2|a+x(1+\sqrt{p})|^2.
\end{equation}
The triangle inequality together with $0\leq p\leq 1$ gives
\begin{equation}
g(p)\leq 1+2\eps^2|a|^2 + 2\eps^2|x|^2(1-\sqrt{p})\leq \big(1+\max(2\eps^2|a|^2,1)\big)\kex^{2}.
\end{equation}
The upper bound $\eps|a|\leq C_a$ and the estimate
$1+\max(2\eps^2|a|^2,1)\leq 2+2C_a^2$ together with a trivial integral gives
that $I$ is bounded from below as
\begin{equation}
I\geq C_2 \kex^{r-2}, \ \text{where}\ C_2:=\frac{1}{2\big(2(1+C_a^2)\big)^{(2-r)/2}}.
\end{equation}
Inserting this result into \eqref{eq:LI} concludes the Lemma.
\end{proof}
Now consider Lemma \ref{lem:c1}. It is a combination of the inequalities 
Theorem 13 and Theorem 19 in \cite{Hardy+Littlewood+Polya}
\begin{lemma}\label{lem:maxmin}
Let $y\in \mathbb{R}$ and $\be\geq 0$ then
\begin{equation}
\frac{1}{2^{\max(0,\frac{2-\be}{2})}} 
\leq\frac{(1+y^2)^{\be/2}}{1+|y|^\be} \leq 2^{\max(0,\frac{\be-2}{2})}.
\end{equation}
\end{lemma}
\begin{proof}
Denote
\begin{equation}
f(y,\be):=\frac{(1+y^2)^{\be/2}}{1+|y|^\be}.
\end{equation}
We first note that $f(y,\be)=f(-y,\be)$, thus we can restrict our
attention to $y\geq 0$. At $y=0$ we have $f(0,\be)=1$ and at $y=\infty$, 
$f(\infty,\be)=1$. The function is differentiable for $y>0$, 
the only critical point for $y>0$ is at $y=1$, where the function
takes the value 
\begin{equation}
f(1,\be)=2^{\be/2-1}.
\end{equation}
If $\be>2$ its a maximum, and if $\be<2$ its a minimum, the lemma
follows.
\end{proof}
To prove Lemma \ref{lem:c3} we begin by stating it again.
\begin{lemma*}
Let $y\in \mathbb{R}$ and $r\geq 2$. Then
\begin{equation}\label{eq:b18}
\left|1+y\right|^r - 1 - r y \geq \frac{1}{2^{r-2}}|y|^r.
\end{equation}
\end{lemma*}
\begin{proof}
Denote 
\begin{equation}
f(y\, ;r):=\left|1+y\right|^r - 1 - r y - \frac{1}{2^{r-2}}|y|^r.
\end{equation}
The lemma is equivalent to $f\geq 0$, for $r\geq 2$. We note that $f$ is
twice differentiable at all points except $y=-1$ and $y=0$. We observe that
the inequality is satisfied for both of these points since
we have $f(0\, ;r)=0$ and $f(-1,r)=r-1-\frac{1}{2^{r-1}}>0$ for $r\geq 2$. 
Consider the derivative of $f$ with respect to $y$:
\begin{equation}\label{eq:df}
\partial_y f(y\, ;r)=r\left(\sign(1+y)|1+y|^{r-1}-\frac{1}{2^{r-2}}\sign(y)|y|^{r-1}-1\right).
\end{equation}
We wish to show that $f$ decays monotonically on $y<0$
and hence, that 
$\partial_y f\leq 0$ for $-1<y<0$ and $y<-1$. 
We also wish to show that $f$ increases monotonically for $y>0$ with
$\partial_y f\geq 0$. To show this, consider first the case $y>0$:
using that $(1+y)^{r-1}\geq 1+ y^{r-1}$, we have $\partial_y f>0$ for $y>0$.

For the interval $-1<y<0$, use that $b^r<b$ for any $b\in(0,1)$, to obtain
\begin{multline}
\partial_y f(y\, ; r)=(1-|y|)^{r-1}+2^{2-r}|y|^{r-1} -1 <
-|y|+2^{2-r}|y| \\ =-\left(1-2^{2-r}\right)|y|\leq 0, \ r\geq 2.
\end{multline}

For the last interval $y< -1$ we re-write \eqref{eq:df} as
\begin{equation}
\partial_y f(y\, ;r)=-r\left(\left(|y|-1\right)^{r-1} -
2^{2-r}|y|^{r-1} + 1\right).
\end{equation}
Upon calculating $\partial_y^2 f$, and solving $\partial_y^2 f=0$ for
$y$ in this interval we find that $\partial_y f$ has a maximum at
$|y|=2$ with value $\partial_y f(2\, ; r)=0$. Hence, $\partial_y f\leq
0$ for $y\leq -1$ and $-1<y<0$.  We have showed that $f$ decays
monotonically on $y<0$ and increases monotonically on $y>0$, and $f(0\,
;r)=0$. Hence $f\geq 0$ for all $y\in\mathbb{R}$, which proves the
lemma.
\end{proof}

\section{Upper bound on $\VR$ and $\nabla_a \VR$}

In this appendix we estimate $\VR$ and derivatives of $\VR$ from above. 
From the proof of 
Lemma \ref{lem:RVbd} we have the following identity
\begin{equation}\label{eq:Vtaylor}
\VR(x)=\int_0^1 (1-\theta)x\cdot \Hess V(a+\theta x)\cdot x\diff \theta.
\end{equation}
Furthermore, in \eqref{eq:Vup} we made the assumptions that, for
$\beta\in\mathbb{Z}^d$ and $|\beta|\leq 3$,
\begin{equation}\label{eq:up}
|\partial_x^\beta V(x)|\leq C_V
\eps^{|\beta|}\kex^{r-|\beta|}.
\end{equation}
We begin with the following result for derivatives of $\VR$.
\begin{lemma}\label{lem:RVup}
  Let $a,x\in \mathbb{R}^d$, $0<\eps\in\mathbb{R}$ and $\eps|a|\leq
  C_a$, as in Corollary~\ref{cor:apest}.  Furthermore let 
  $\beta\in \mathbb{Z}^d$, with $0\leq \beta_j \leq 1$
  $\forall j=1,...,d$ and $|\beta| =  1$ Then, (i) if $r\geq 2$:
\begin{equation}
|\partial^\beta_a \VR| \leq C_1 \eps^{3}
|x|^2\kex^{\max(r-3,0)},
\end{equation}
or (ii) if $1\leq r<2$:
\begin{equation}
|\partial^\beta_a  \VR| \leq C_2 
\eps^{3}|x|^2\kex^{r-2},
\end{equation}
where
\begin{equation}
C_1:=2^{-1}C_Vd(2(1+C_a^2))^{\max(r-3,0)/2}, \ \ 
C_2:=C_Vd\big(6\sqrt{2}+\ln(1+C_a)\big).
\end{equation}
Here $C_V$ is the constant in \eqref{eq:Vup}.
\end{lemma}
\begin{corollary}\label{cor:RVup}
Under the same conditions as in Lemma \ref{lem:RVup}
we have
\begin{equation}
|\partial_a^\beta \VR|\leq C\eps^{3}|x|^2\kex^{r-2},
\end{equation}
where $C$ depends on $C_1$ and $C_2$ above.
\end{corollary}
\begin{proof}
Use that $\kex^{\max(r-3,0)}\leq \kex^{r-2}$ in Lemma \ref{lem:RVup}.
\end{proof}
\begin{proof}[Proof of Lemma \ref{lem:RVup}]
  For the case $x=0$, $(\partial^\beta_a \VR)(0)=0$ 
and thus the Lemma is
  satisfied.  For $x\neq 0$, and since $V\in\C{3}$, we have from
  \eqref{eq:Vtaylor} that 
\begin{equation}
\partial_{a}^\beta \VR(x,t)=\sum_{k,l=1}^d x_lx_k \int_0^1 (1-\theta)
(\partial_{x}^\beta\partial_{x_k}\partial_{x_l} V)(a+\theta x) \diff \theta.
\end{equation}
The upper bound on the potential, \eqref{eq:up}, gives
\begin{equation}
|\partial_{a_j} \VR(x)|\leq \eps^{3} C_V 
\sum_{k,l=1}^d |x_lx_k| 
\int_0^1 (1-\theta) (1+\eps^2|a+\theta x|^2)^{(r-3)/2} 
\diff \theta,
\end{equation}
here $C_V$ is the constant in \eqref{eq:Vup}. 
To estimate $|\partial_a^\beta
\VR|$,  we use the inequality 
\begin{equation}
\sum_{k,l=1}^d |x_kx_l|\leq d|x|^2,
\end{equation}
to obtain 
\begin{equation}
|\partial_a^\beta \VR(x)|\leq C_V d \eps^{3}|x|^2 \int_0^1 (1-\theta) 
(1+\eps^2|a+\theta x|^2)^{(r-3)/2}  =d C_V \eps^{3} |x|^2 I.
\end{equation}

To estimate the integrand, we consider first case (i), with 
$r>3$, $\eps|a|\leq C_a$. Before we estimate the integral $I$,
we estimate part of its integrand with 
the triangle inequality together with $\eps|a|\leq C_a$ and 
$\theta\leq 1$ to obtain
\begin{equation}
1+\eps^2 |a+\theta x|^2\leq 1+ 2C_a^2 + 2|x|^2 \leq 
(1+\max(2C_a^2,1))\kex^2.
\end{equation}
Thus 
\begin{equation}\label{eq:uest}
|\partial_a^\beta \VR(x,t)|\leq \tilde{C}_1 \eps^{3}|x|^2\kex^{r-3},
\end{equation}
where
\begin{equation}\label{eq:const1}
\tilde{C}_1:=2^{-1}C_V d(2(1+C_a^2))^{(r-3)/2}.
\end{equation}

To extend this case to include $r\geq 2$, we note that for $r\in[2,3]$
the exponent in the integrand of I,  $r-3\leq 0$, 
and that $1+\eps^2|a+\theta x|^2\geq 1$ to obtain
\begin{equation}
I \leq 2^{-1}.
\end{equation}
We conclude that for $r\in[2,3]$
$|\partial_a^\beta \VR|
\leq 2^{-1}dC_V\eps^{3}|x|^2$. Thus upon changing
\eqref{eq:uest} and \eqref{eq:const1} into
\begin{equation}
|\partial_a^\beta \VR(x)|\leq C_1 \eps^{3}
|x|^2\kex^{\max(r-3,0)},
\end{equation}
where
\begin{equation}
C_1:=2^{-1}C_V d(2(1+C_a^2))^{\max(r-3,0)/2},
\end{equation}
part (i) is proved.

For the case (ii), with $r< 2$ we need a more precise estimate that the case of 
$r\in[2,3]$. To obtain this, recall that $x\neq 0$ and let us
introduce the notations $\hat{x}:=x/|x|$, $\ap:=a\cdot \hat{x}$ and
$\ab:=a-\ap$. Then $|a+\theta x|^2 = \ab^2+\big|\ap+\theta
|x|\big|^2$.  By the change of variables $y=\eps(\ap+\theta |x|)$ and
that $1+\ab^2\geq 1$, we find
\begin{equation}
I\leq \frac{1}{\eps|x|}\int_{\eps \ap}^{\eps(\ap+|x|)} 
\frac{\diff y}{(1+y^2)^{(3-r)/2}}=:\frac{1}{\eps|x|}I_2.
\end{equation}
The goal is to show that $\kex^{2-r} I$ is bounded by a constant independent
of $\eps$. To show this, we consider two intervals for $|x|$
first $\eps |x|\leq 1$.
For this interval $\kex^{2-r}\leq 2^{2-r}$ 
and $I_2 \leq \eps|x|$, thus $\kex^{2-r} I\leq 2$.

For the intervals $\eps|x|\geq 1$ and $1\leq r\leq 1.5$. We show that $I_2$ is
bounded by a constant. Indeed, regardless of the values of $\ap$ and $|x|$
we have
\begin{equation}
I_2 \leq 2\int_0^\infty \frac{\diff y}{(1+y^2)^{(3-r)/2}}=
\sqrt{\pi}\frac{\Gamma(1-\frac{r}{2})}{\Gamma(\frac{r-3}{2})}\leq 6
\end{equation}
for $1\leq r\leq 1.5$. Thus 
\begin{equation}
\kex^{2-r}I\leq 6\frac{\kex^{2-r}}{\eps |x|}\leq 6\sqrt{2}.
\end{equation}

For $r\in [1.5,2]$ we need a
better estimate, we use that $(1+y^2)^{(3-r)/2}\geq (1+y^2)^{1/2}$,
thus
\begin{equation}
I_2 \leq \int_{\eps\ap}^{\eps(\ap+|x|)} \frac{\diff y}{\sqrt{1+y^2}} =
 \ln(\frac{\eps\ap+\eps|x|+\sqrt{1+\eps^2|a+x|^2}}
     {\eps\ap+\sqrt{1+\eps^2\ap^2}}).
\end{equation}
To estimate this, we consider four different regions, For $\ap>0$ and
$|x|>|a|$ it is bounded by $\ln(1+4\eps|x|)$.  For $\ap>0$ and
$|x|\leq |a|$ it is bounded by $\ln 2$.  For $\ap<0$ and
$|x|<|a|$ it is bounded by $\ln(1+2C_a)$. For $\ap<0$ and
$|x|>|a|$ it is bounded by $\ln\big((1+2C_a)(1+4\eps|x|)\big)$.  Thus
\begin{equation}\label{eq:IntC2}
I_2 \leq \ln(2+2C_a)+\ln(1+4\eps |x|)\leq (\eps |x|)^{1/2}\ln \big(10(1+C_a)\big), \ \text{for}\ \eps|x|\geq 1.
\end{equation}
where we have used that for $\eps|x|\geq 1$, $q\geq 0$ we have
$q+\ln(1+4\eps|x|)\leq (q+\ln 5)(\eps|x|)^{1/2}$.
Thus 
\begin{equation}
\kex^{2-r}I\leq \ln(10(1+C_a))\frac{\kex^{2-r}}{(\eps|x|)^{1/2}} \leq
2^{1/4}\ln(10(1+C_a)).  
\end{equation}

Comparing the constants above for the $I$ estimate we find that
\begin{equation}
C_2:=C_Vd(6\sqrt{2}+\ln(1+C_a)),
\end{equation}
is sufficient. This concludes the proof of the lemma. 
\end{proof}

To bound $\VR$ from above we could use the same methods as above, but
the upper bound will be to large to fit into the energy space. But
we have the following
\begin{lemma}\label{lem:VRup2}
For $r\geq 1$ and $\eps|a|\leq C_a$
\begin{equation}
\VR\leq C_1 (1+\eps^2|x|^2\kex^{r-2}),
\end{equation}
where
\begin{equation}
C_1:=2C_V(2+2C_a^2)^{(r-1)/2}.
\end{equation}
\end{lemma}
\begin{proof}
For $r\geq 2$, we there exists a $\theta\in[0,1]$ such that
\begin{equation}
\VR\leq C_V\eps^2|x|^2\kax{x\theta+a}^{r-2}.
\end{equation}
Since $r-2\geq 0$ we estimate $\eps|x\theta+a|\leq \eps|x|+C_a$ and
$(1+2a|x|^2+2C_a^2)\leq (2+2C_a^2)\kex^2$ we obtain the lemma for
$r\geq 2$ as
\begin{equation}\label{eq:Cbd1}
\VR\leq C_V(2+2C_a^2)^{(r-2)/2}|\eps|^2|x|^2\kex^{r-2}.
\end{equation} 
For $r\in [1,2)$ we use that there exists a $\theta\in [0,1]$ such that 
\begin{equation}
\VR=(\nabla V(x\theta+a,t)-\nabla V(a))\cdot x\leq 
C_V\eps|x|(\kax{\eps(x\theta+a)}^{r-1}+\kax{C_a}^{r-1}).
\end{equation}
Once again $\eps|x\theta+a|\leq \eps|x|+C_a$ and we obtain
\begin{equation}\label{eq:CC1}
\VR\leq 
C_V\eps|x|\big((2+2C_a^2)^{(r-1)/2}\kax{\eps x}^{r-1}+
\kax{C_a}^{r-1}\big).
\end{equation}
To estimate the second term, recall \eqref{eq:gex}, to get
\begin{equation}
\eps|x|\leq 2^{-1}+(\eps|x|)^2\kex^{r-2}.
\end{equation}
To estimate the first term in \eqref{eq:CC1}, let $y=\eps|x|\geq 0$, 
$r\in[1,2]$ and calculate
\begin{equation}
y\kax{y}^{r-1}-y^2\kax{y}^{r-2} = 
y\kax{y}^{r-1}(1-\frac{y}{\kax{y}})=
\frac{y\kax{y}^{r-1}}{(y+\kax{y})\kax{y}}\leq 
\frac{y}{(y+\kax{y})}\leq \frac{1}{2}. 
\end{equation} 
Thus
\begin{equation}
\eps|x|\kex^{r-1}\leq 2^{-1}+(\eps|x|)^2\kex^{r-2}.
\end{equation}
Collecting the above two terms gives 
\begin{equation}\label{eq:Cbd2}
\VR\leq 2
C_V(2+2C_a^2)^{(r-1)/2}(1+(\eps|x|)^2\kax{\eps x}^{r-2}).
\end{equation}
Since \eqref{eq:Cbd2} for $r\geq 2$ is larger than \eqref{eq:Cbd1} we
have proved the lemma.
\end{proof}

\section{Bound in energy-space}\label{sec:ene}

In \Eref{eq:1012} we show that 
\begin{equation}
\|w\|_{0} := \nrmHo{w}+\Bnrm{\eps|x|\kex^{(r-2)/2}w}\leq C\epsT.
\end{equation}
We want to show that $\|w\|_{\Espace}\leq C_r \|w\|_0$.
This result follows from the following lemma:
\begin{lemma}\label{lem:eqv}
For $r\in(0,\infty)$ there exists a constant $c_r$ such that 
\begin{equation}\label{eq:no1}
0<1+\min(0,c_r) \leq \frac{1+y^2\kax{y}^{r-2}}{\kax{y}^r}\leq 1+\max(0,c_r)<2,
\end{equation}
where
\begin{equation}
c_r:=\frac{2-r}{2}\left(\frac{2}{r}\right)^{2/(r-2)}, \ r\neq 2,
\end{equation}
and for $r=2$, $c_r=0$.
\end{lemma}
\begin{proof}
Denote
\begin{equation}
f(z)=\frac{1+(z^2-1)z^{r-2}}{z^r}=1+z^{-r}-z^{-2}, \ z\geq 1.
\end{equation}
Note that for $z^2=1+y^2$, $f$ is the function we want to estimate for
the lemma. For $r=2$, $f=1$, thus $c_r=0$.  The function $f$ is at
least $\C{1}$ for $z\geq 1$. Now consider $r\neq 2$. We note that
$f(1)=1$ and $f\rightarrow 1$ as $z\rightarrow \infty$.  
The critical point on $[1,\infty)$ of $f$ is at $z_c:=(r/2)^{1/(r-2)}>1$,
where the function take the value
\begin{equation}
f(z_c)=1+\frac{2-r}{r}\left(\frac{2}{r}\right)^{2/(r-2)}.
\end{equation}
A maximum(minimum) for $r<2$($r>2$). This concludes the
proof.
\end{proof}

\section{A family of time-dependent solutions}\label{sec:fam}

In this appendix, we construct a family of solutions to the
nonlinear Schr\"odinger equation with a quadratic, time-independent
potential.

Let $\psi(x,t)$ have the form
\begin{equation}
\psi(x,t)=\lexp{\iu \mom(t)\cdot(x-a(t))+\iu \gamma(t)}\QQ(x-a(t)),
\end{equation}
where $\QQ$ is a real-valued function, not yet determined.
We substitute this function into \eqref{eq:NLS}, and let $y:=x-a$ to obtain
\begin{equation}
0= \dot\mom\cdot y\QQ+ (\dot\gamma +
\mom^2-\dot a\cdot p)\QQ +\iu \nabla \QQ\cdot(\dot a -2\mom) 
-\Laplace \QQ -f(\QQ)+V(y+a)\QQ. 
\end{equation}
By adding and subtracting the terms $(\freq+ V(a))\QQ$ and $\nabla
V(a)\cdot y\QQ$ and as usual defining $\VR:=V(y+a)-V(a)-\nabla V(a)\cdot
y$ we find
\begin{multline}\label{eq:solV}
0= (\dot\mom+\nabla V(a))\cdot y\QQ + (\dot\gamma +
\mom^2-\dot a\cdot \mom +V(a)-\freq)\QQ +\iu \nabla \QQ\cdot(\dot a -2\mom) 
\\ +(-\Laplace+\freq) \QQ -f(\QQ)+\VR\QQ. 
\end{multline}
If we choose
\begin{equation}
\dot\mom = -\nabla V(a), \ \dot a=2\mom,\ \dot\gamma=\mom^2+\freq-V(a),
\end{equation}
then the Eqn.~\ref{eq:solV} reduces to
\begin{equation}\label{eq:Vsol}
0= -\Laplace\QQ +\freq \QQ -f(\QQ)+\VR\QQ,
\end{equation}
where $\QQ=\QQ(y)$, and $\Laplace=\sum \partial_{y_j}^2$.
In general this equation is time-dependent due to the appearance of $a$
in $\VR$, but for potentials of the form 
$V(x)=x\cdot A\cdot x+v\cdot x + d$, with constant
matrix $A$, vector $v$ and scalar $d$, we have
\begin{multline}
\VR = (y+a)\cdot A\cdot (y+a) + v\cdot (y+a)+d
- (a\cdot A\cdot a + v\cdot a +d) \\ - (a\cdot A\cdot y
+y\cdot A \cdot a + v\cdot y) = y\cdot A\cdot y.
\end{multline}
The right-hand side is independent of $a$, and hence of time. 
Equation~\eqref{eq:solV} reduces to
\begin{equation}\label{eq:Vsol2}
0= -\Laplace\QQ +\freq \QQ -f(\QQ)+y\cdot A\cdot y \QQ.
\end{equation}
Thus, if there exists nontrivial solutions to this equation, we have
constructed a family of solutions $\lexp{\iu\mom(t)\cdot
  (x-a)+\iu\gamma}\QQ(x-a)$, where $\QQ$ solves \eqref{eq:Vsol2}.
Existence of solutions to a general class of equations that includes
\eqref{eq:Vsol2} under some restrictions on $b:=\freq + y\cdot A\cdot
y$ and with a class of local nonlinearities is shown by
Rabinowitz~\cite{Rabinowitz1992} and extended to more general
potentials by Sikarov~\cite{Sirakov00}.  Sikarov require the following
potential conditions: $b>-c$, where $|c|<\infty$,
\begin{equation}
\inf_{u\in\Hone,\nrm{u}=1} \nrm{\nabla u}^2+\dotp{bu}{u}>0
\end{equation}
and that $b$ grows to infinity in almost all directions as
$|y|\rightarrow \infty$.

\bigskip

\noindent
{\it{\bf Acknowledgement:}} part of this work was done while one of
the authors (IMS) was visiting ETH Z\"{u}rich and ESI Vienna.
IMS is grateful to J. Fr\"ohlich and to P.C. Aichelburg and P. Bizo\'n
for their hospitality.


\begin{thebibliography}{10}

\bibitem{Berestycki+LionsI1983}
H.~Berestycki and P.-L. Lions.
\newblock Nonlinear scalar field equations. {I}. {E}xistence of a ground state.
\newblock {\em Arch. Rational Mech. Anal.}, 82(4):313--345, 1983,
  ams:MR0695535.

\bibitem{Berestycki+LionsII1983}
H.~Berestycki and P.-L. Lions.
\newblock Nonlinear scalar field equations. {II. E}xistence of infinitely many
  solutions.
\newblock {\em Arch. Rational Mech. Anal.}, 82(4):347--375, 1983,
  ams:MR0695536.

\bibitem{Berestycki+Lions+Peletier1981}
H.~Berestycki, P.-L. Lions, and L.~A. Peletier.
\newblock An {ODE} approach to the existence of positive solutions for
  semilinear problems in {$\mathbb{R}^{N}$}.
\newblock {\em Indiana Univ. Math. J.}, 30(1):141--157, 1981, ams:MR0600039.

\bibitem{Bronski+Jerrard2000}
J.~C. Bronski and R.~L. Jerrard.
\newblock Soliton dynamics in a potential.
\newblock {\em Math. Res. Lett.}, 7(2-3):329--342, 2000, ams:MR1764326.

\bibitem{BP92}
V.~S. Buslaev and G.~S. Perel'man.
\newblock Scattering for the nonlinear {S}chr{\"o}dinger equation: states that
  are close to a soliton.
\newblock {\em Algebra i Analiz}, 4(6):63--102, 1992.

\bibitem{Buslaev+Perelman1995}
V.~S. Buslaev and G.~S. Perel'man.
\newblock On the stability of solitary waves for nonlinear {S}chr\"{o}dinger
  equations.
\newblock {\em Amer. Math. Soc. Transl. Ser.}, 2(164):74--98, 1995.

\bibitem{Buslaev+Sulem2002}
V.~S. Buslaev and C.~Sulem.
\newblock On asymptotic stability of solitary waves for nonlinear
  {S}chr{\"o}dinger equations.
\newblock {\em Ann. IHP. Analyse Nonlin\'eaire}, 20:419--475, 2003,
  doi:10.1016/S0294-1449(02)00018-5.

\bibitem{Carles2003}
R.~Carles.
\newblock Semi-classical {S}chr\"odinger equations with harmonic potential and
  nonlinear perturbation.
\newblock {\em Ann. I.H.P., Analyse non lin\'eaire}, 20(3):501--542, 2003,
  doi:10.1016/S0294-1449(02)00027-6.

\bibitem{Cazenave1996}
T.~Cazenave.
\newblock {\em An introduction to nonlinear {S}chr\"{o}dinger equations}.
\newblock Number~26 in Textos de M\'etodos Matem\'aticos. Instituto de
  Matematica - UFRJ, Rio de Janeiro, RJ, third edition, 1996.

\bibitem{Cuccagna2001}
S.~Cuccagna.
\newblock Stabilization of solutions to nonlinear {S}chr\"{o}dinger equations.
\newblock {\em Comm. Pure Appl. Math.}, 54(9):1110--1145, 2001,
  doi:10.1002/cpa.1018.

\bibitem{Cuccagna2002}
S.~Cuccagna.
\newblock Asymptotic stability of the ground states of the nonlinear
  {S}chr{\"o}dinger equation.
\newblock {\em Rend. Istit. Mat. Univ. Trieste}, 32(suppl. 1):105--118, 2002.

\bibitem{FGJS-I}
J.~Fr\"ohlich, S.~Gustafson, B.~L.~G. Jonsson, and I.~M. Sigal.
\newblock Solitary wave dynamics in an external potential.
\newblock {\em Comm. Math. Phys.}, 250(3):613--642, 2004,
  doi:10.1007/s00220-004-1128-1.

\bibitem{Frohlich+Tsai+Yau2000}
J.~Fr\"{o}hlich, T.-P. Tsai, and H.-T. Yau.
\newblock On a classical limit of quantum theory and the non-linear {H}artree
  equation.
\newblock {\em Geom. Funct. Anal.}, Special Volume, Part I:57--78, 2000,
  ams:MR1826249.

\bibitem{Frohlich+Tsai+Yau2002}
J.~Fr\"{o}hlich, T.-P. Tsai, and H.-T. Yau.
\newblock On the point-particle ({N}ewtonian) limit of the non-linear {H}artree
  equation.
\newblock {\em Comm. Math. Phys.}, 225(2):223--274, 2002,
  doi:10.1007/s002200100579.

\bibitem{Gang+Sigal2004}
Z.~Gang and I.~M. Sigal.
\newblock Asymptotic stability of nonlinear {S}chr\"odinger equations with
  potential.
\newblock preprint, 2004, ArXiv: math-ph/0411050.

\bibitem{Grillakis+Shatah+Strauss1990}
M.~Grillakis, J.~Shatah, and W.~Strauss.
\newblock Stability theory of solitary waves in the presence of symmetry. {II}.
\newblock {\em J. Funct. Anal.}, 94(2):308--348, 1990,
  doi:10.1016/0022-1236(90)90016-E.

\bibitem{Gustafson+Nakanishi+Tsai2004}
S.~Gustafson, K.~Nakanishi, and T.-P. Tsai.
\newblock Asymptotic stability and completeness in the energy space for
  nonlinear {S}chr{\"o}dinger equations with small solitary waves.
\newblock {\em Int. Math. Res. Not.}, 2004(66):3559--3584, 2004, doi:
  10.1155/S1073792804132340.

\bibitem{Hardy+Littlewood+Polya}
G.~H. Hardy, J.~E. Littlewood, and G.~P\'olya.
\newblock {\em Inequalities}.
\newblock Cambridge University Press, second edition, 1967.

\bibitem{Keraani2002}
S.~Keraani.
\newblock Semiclassical limit of a class of {S}chr\"odinger equations with
  potential.
\newblock {\em Comm. PDE.}, 27(3 \& 4):693--704, 2002,
  doi:10.1081/PDE-120002870.

\bibitem{Enno}
E.~Lenzmann.
\newblock Global well-posedness for {NLS} with external potentials.
\newblock Private communication, part of upcoming thesis, 2005.

\bibitem{McLeod1993}
K.~Mc{L}eod.
\newblock Uniqueness of positive radial solutions of {$\Laplace u + f(u)=0$} in
  {$\mathbb{R}^n$}, {II}.
\newblock {\em Am. Math. Soc.}, 339(2):495--505, 1993, ams:MR1201323.

\bibitem{Perelman2004}
G.~Perelman.
\newblock Asymptotic stability of multi-soliton solutions for nonlinear
  {S}chr\"o\-dinger equations.
\newblock {\em Comm. PDE}, 29(7-8):1051--1095, 2004,
  \\doi:10.1081/PDE-200033754.

\bibitem{Rabinowitz1992}
P.~H. Rabinowitz.
\newblock On a class of nonlinear {S}chr\"odinger equations.
\newblock {\em Z. Angew. Math. Phys.}, 43(2):270--291, 1992, ams:MR1162728.

\bibitem{RSS}
I.~Rodnianski, W.~Schlag, and A.~Soffer.
\newblock Asymptotic stability of $n$-soliton states of {NLS}.
\newblock To appear in Comm. Pure and Appl. Math., 2003, ArXiv:math.AP/0309114.

\bibitem{Sirakov00}
B.~Sirakov.
\newblock Existence and multiplicity of solutions of semi-linear elliptic
  equations in $\mathbb{R}^n$.
\newblock {\em Calc. Var.}, 11(2):119--142, 2000, doi: 10.1007/s005260000010.

\bibitem{Soffer+Weinstein1988}
A.~Soffer and M.~I. Weinstein.
\newblock Multichannel nonlinear scattering theory for nonintegrable equations.
\newblock In {\em Integrable systems and applications (\^Ile d'Ol\'eron,
  1988)}, volume 342 of {\em Lecture Notes in Phys.}, pages 312--327. Springer,
  Berlin, 1989, ams:MR1034551.

\bibitem{Soffer+Weinstein1990}
A.~Soffer and M.~I. Weinstein.
\newblock Multichannel nonlinear scattering for nonintegrable equations.
\newblock {\em Comm. Math. Phys.}, 133(1):119--146, 1990, ams:MR1071238.

\bibitem{Soffer+Weinstein1992}
A.~Soffer and M.~I. Weinstein.
\newblock Multichannel nonlinear scattering for non{\-}integrable equations
  {II}. {T}he case of anisotropic potentials and data.
\newblock {\em J. Differential Equations}, 98(2):376--390, 1992,
  doi:10.1016/0022-0396(92)90098-8.

\bibitem{Soffer+Weinstein2004}
A.~Soffer and M.~I. Weinstein.
\newblock Selection of the ground state for nonlinear {S}chr\"{o}dinger
  equations.
\newblock {\em Reviews in Math. Phys.}, 16(8):977--1071, 2004,
  \\doi:10.1142/S0129055X04002175.

\bibitem{Tsai+Yau2002}
T.-P. Tsai and H.-T. Yau.
\newblock Asymptotic dynamics of nonlinear {S}chr\"{o}dinger equations:
  resonance-dominated and dispersion-dominated solutions.
\newblock {\em Comm. Pure Appl. Math.}, 55(2):153--216, 2002,
  doi:10.1002/cpa.3012.

\bibitem{Tsai+Yau2002b}
T.-P. Tsai and H.-T. Yau.
\newblock Relaxation of excited states in nonlinear {S}chr\"odinger equations.
\newblock {\em Int. Math. Res. Not.}, 2002(31):1629--1673, 2002,
  \\doi:10.1155/S1073792802201063.

\bibitem{Tsai+Yau2002c}
T.-P. Tsai and H.-T. Yau.
\newblock Stable directions for excited states of nonlinear {S}chr{\"o}dinger
  equations.
\newblock {\em Comm. PDE}, 27(11-12):2363--2402, 2002,
  doi:10.1081/PDE-120016161.

\bibitem{Weinstein1985}
M.~I. Weinstein.
\newblock Modulational stability of ground states of nonlinear
  {S}chr\"{o}dinger equations.
\newblock {\em SIAM J. Math. Anal.}, 16(3):472--491, 1985, ams:MR0783974.

\bibitem{Yajima+Zhang2001}
K.~Yajima and G.~Zhang.
\newblock Smoothing property for {S}chr\"odinger equations with potential
  superquadratic at infinity.
\newblock {\em Comm. Math. Phys.}, 221(3):573--590, 2001,
  doi:10.1007/s002200100483.

\bibitem{Yajima+Zhang2004}
K.~Yajima and G.~Zhang.
\newblock Local smoothing property and {S}tri\-ch\-artz inequality for
  {S}chr\"odinger equations with potentials super\-quadratic at infinity.
\newblock {\em J. Differential Equations}, 202(1):81--110, 2004,
  doi:10.1016/j.jde.2004.03.027.

\end{thebibliography}

\end{document}